\documentclass[fleqn,usenatbib]{mnras}

\usepackage{verbatim}
\usepackage{pdfpages}
\usepackage{float}
\usepackage{tabularx}
\usepackage{multirow}
\usepackage[normalem]{ulem}
\usepackage[displaymath, mathlines, pagewise]{lineno}

\restylefloat{table}
\usepackage[T1]{fontenc}

\DeclareRobustCommand{\VAN}[3]{#2}
\let\VANthebibliography\thebibliography
\def\thebibliography{\DeclareRobustCommand{\VAN}[3]{##3}\VANthebibliography}

\usepackage{subcaption}
\usepackage{graphicx}	% Including figure files
\usepackage{amsmath}	% Advanced maths commands
\usepackage{amssymb}	% Extra maths symbols
\usepackage{enumitem}   % enumerate specialties
\usepackage{newtxtext,newtxmath} %adjusted order
\newcommand{\DG}[1]{\textcolor{red}{DG: #1}}

\newcommand{\revtwo}[1]{{\textcolor{black}{#1}}}

\newcommand{\rev}[1]{{#1}} %comment out if wanting revisions visible

\title[DESI Calibration of the Color-Redshift Relation]{DESI Complete Calibration of the Color-Redshift Relation (DC3R2): Results from early DESI data}

\author[J. McCullough et al.]{
J.~McCullough,$^{1,2,3}$\thanks{E-mail: jmccull@stanford.edu}
D.~Gruen,$^{3,4}$
A.~Amon,$^{5,6}$
A.~Roodman,$^{1,2}$
D.~Masters,${^7}$
A.~Raichoor,${^8}$
D.~Schlegel,${^8}$\newauthor
R.~Canning,${^9}$ 
F.~J.~Castander,$^{10,11}$
J.~DeRose,${^8}$
R.~Miquel,$^{12,13}$
J.~Myles,$^{1,2}$
J.~A.~Newman,$^{14}$
A.~Slosar,$^{15}$\newauthor
J.~Speagle,$^{16}$
M.~J.~Wilson,$^{17}$
J.~Aguilar,$^{8}$ %lbl
S.~Ahlen,$^{18}$ %
S.~Bailey,$^{8}$ %
D.~Brooks,$^{19}$ %
T.~Claybaugh,$^{8}$ %
S.~Cole,$^{17}$ \newauthor%
K.~Dawson,$^{20}$ 
A.~de la Macorra,$^{21}$ %
P.~Doel,$^{19}$ %
J.~E.~Forero-Romero,$^{22}$ %
S.~Gontcho A Gontcho,$^{8}$
J.~Guy,$^{8}$ \newauthor %
R.~Kehoe,$^{23}$ %
A.~Kremin,$^{8}$
M.~Landriau,$^{8}$ %
L.~Le~Guillou,$^{24}$ %
M.~Levi,$^{8}$ %
M.~Manera,$^{13}$ %
P.~Martini,$^{25}$
A.~Meisner,$^{26}$ \newauthor %
J.~Moustakas,$^{27}$ %
J.~Nie,$^{28}$
W.~J.~Percival,$^{29}$ %
C.~Poppett,$^{8,30}$ %
F.~Prada,$^{31}$ %
M.~Rezaie,$^{32}$ %
G.~Rossi,$^{33}$ %
E.~Sanchez,$^{34}$ \newauthor %%
H.~Seo,$^{35}$ %
G.~Tarl\'{e},$^{36}$
B.~A.~Weaver,$^{26}$ %
Z.~Zhou,$^{37}$ %
H.~Zou,$^{37}$ %
\textit{DESI Collaboration}
}

\date{Accepted 2024 May 9. Received 2024 May 9; in original form 2023 September 12}

\def\today{3 June 2024}

\pubyear{2024}

\begin{document}
\label{firstpage}
\pagerange{\pageref{firstpage}--\pageref{lastpage}}
\maketitle
\newcommand{\note}
  [ 1 ]{
  {\color{black}#1\color{black} } }
\begin{abstract}
We present initial results from the Dark Energy Spectroscopic Instrument (DESI) Complete Calibration of the Color-Redshift Relation (DC3R2) secondary target survey. Our analysis uses \note{230k} galaxies that overlap with KiDS-VIKING \textit{ugriZYJHK$_s$} photometry to calibrate the color-redshift relation and to inform photometric redshift (photo-\textit{z}) inference methods of future weak lensing surveys. Together with Emission Line Galaxies (ELGs), Luminous Red Galaxies (LRGs), and the Bright Galaxy Survey (BGS) that provide samples of complementary color, the DC3R2 targets help DESI to span \note{56\%} of the color space visible to Euclid and LSST with high confidence spectroscopic redshifts. The effects of spectroscopic completeness and quality are explored, as well as systematic uncertainties introduced with the use of common Self Organizing Maps trained on different photometry than the analysis sample. We further examine the dependence of redshift on magnitude at fixed color, important for the use of bright galaxy spectra to calibrate redshifts in a fainter photometric galaxy sample. We find that noise in the KiDS-VIKING photometry introduces a dominant, apparent magnitude dependence of redshift at fixed color, which indicates a need for carefully chosen deep drilling fields, and survey simulation to model this effect for future weak lensing surveys.
\end{abstract}


%%%%%%%%%%%%%%%%%%%%%%%%%%%%%%%%%%%%%%%%%%%%%%%%%%

%%%%%%%%%%%%%%%%% BODY OF PAPER %%%%%%%%%%%%%%%%%%

\section{Introduction}

\par{Modern cosmology relies on our ability to observe galaxies as tracers of the structure formation and expansion of the Universe. To do this we must map their on-sky positions and, crucially, their positions in all three dimensions. 
%Obtaining distances to galaxies has been historically difficult. However, 
Measuring the redshift, \textit{z}, for galaxies outside of our local group is a good proxy for this third dimension, because the expansion of the universe reddens the light from a galaxy in a way that is monotonic with distance. The most accurate way to measure a galaxy's redshift is via the detection of prominent emission or absorption features with sufficient signal-to-noise ratio in the spectral energy distribution (SED). With this, the observed wavelengths of these features are compared to their known rest frame wavelengths to provide a redshift measurement.  Spectroscopic surveys obtain these many-wavelength observations very successfully via slit masks on the telescope (e.g. on Keck, \citealt{keck_LRS}), and more recently with integral field units (IFUs) (e.g. the Hobby Eberly Telescope Dark Energy Experiment, \citealt{hetdex}), and massively multiplexed instruments with independent optical fibers capable of taking many spectra at once (e.g. DESI, \citealt{DESI_2014, desi_instrument}). The majority of galaxies in our Universe are relatively faint and distant, and low-throughput galaxies are more feasibly observed photometrically -- in optical, near-infrared, and infrared filters -- rather than spectroscopically, due to limits in exposure time. Imaging surveys estimate the distances to galaxies from their colors, measured as the ratio of photometric fluxes in different bandpass filters. Although photometric redshifts are more readily attainable, these estimates are less accurate than spectroscopic redshifts.}

Cosmological measurements from wide imaging surveys, like weak gravitational lensing or galaxy clustering, rely on geometric information. The most recent imaging surveys, such as the Dark Energy Survey (DES, \citealt{desy3_photom}), Subaru Hyper Suprime-Cam (HSC, \citealt{hsc_dr3}) and the Kilo-Degree Survey (KiDS, \citealt{KiDS_DR4}), span a significant fraction of the sky. Accurate estimates of the redshifts of these galaxy samples based on limited information  (e.g. photometry rather than spectroscopy) are required to obtain unbiased cosmological constraints. Indeed, one of the foremost difficulties facing imaging surveys for cosmology lies in calibrating the redshift probability distribution \citep{Myles_2021}. Typically, redshift distributions are estimated and calibrated for an ensemble of galaxies, $n(z)$, and the uncertainty is modelled as an error on the mean redshift of the distribution in the cosmological analysis--as cosmological parameters are most sensitive to shifts in the mean-\textit{z} \citep[see e.g.][]{ Amon_2022, hscy3_real, hscy3_fourier, van_den_Busch_2022}. Calibrating the redshift distribution for an entire ensemble has unique difficulties compared to doing the same for individual galaxies, though individual redshifts have a multitude of other science applications. As ensemble redshift distributions are of the foremost interest to weak lensing cosmology, ensemble calibration is the focus of this paper. 

%One of the foremost difficulties in wide imaging surveys for cosmology lies in calibrating the redshift distribution, $n(z)$, of samples of galaxies selected using limited information. Ensemble redshift distributions directly enter the measurements of cosmological parameters like $\Omega_m$ and $\sigma_8$, and therefore their estimation is. For galaxies with spectral energy distributions (SEDs) observed at high resolution and sufficient signal-to-noise, obtaining a redshift is simply the result of detecting prominent emission or absorption features and comparing to the known rest frame wavelengths. However, faint galaxies are mostly observed photometrically due to limits in exposure time. 

With observations of fluxes in only a few broad bands, the underlying challenge for determining an accurate redshift is a degeneracy between a galaxy's spectral phenotype and redshift (see \citealt{GruenNewman} for a review). A variety of approaches have historically been used to determine galaxy redshifts. Quiescent elliptical galaxies have a consistent drop in light emission at 4000\AA, which enables accurate photo-\textit{z}s with this so-called red sequence of galaxies, which allows the bounding of the redshift between different band passes and is particularly useful for identifying cluster members and lensing galaxy samples (e.g. the RedMaPPer algorithm in \citealt{Rykoff_2014, Rykoff_2016}). While early-type, passive galaxies benefit from having very similar SEDs, this is not necessarily true for many other galaxy types. Template fitting methods can rely on either spectroscopically informed templates or semi-analytic models that are typically constructed with stellar populations (e.g. \citealt{eazy}). An empirical variation of this is a Principal Component Analysis (PCA), but in essence both methods fit the data to a linear combination of templates (see \citealt{salvato2018flavours} for a review). Regardless of the method for redshift estimation, redshift-type degeneracy is an irreducible problem when working with photometry (see discussion of methods that break age-mass-redshift degeneracy in e.g. \citealt{Wang_2023}).

A correct model for the galaxy population (i.e. the mix of templates and their luminosity functions at each redshift, or a large, fully representative reference sample of galaxies with known spectroscopic redshift) would be required to determine correct $n(z)$ for photometric samples despite redshift-type degeneracies. Both of these solutions at present appear unfeasible, though gains in forward modeling the distribution have been made (e.g. \citealt{Alsing_2023}). The issue can be greatly reduced by observing in additional bands that break degeneracies, which motivates a filter set that goes beyond the standard optical broad bands \citep{Buchs_2019, Wright_2019}. With this increased wavelength coverage, however, the color space that the photometric observations occupy becomes high-dimensional. The challenge is to associate like-spectroscopic galaxies with photometric galaxies efficiently across that high-dimensional space. 

Self Organizing Maps (SOMs,\citealt{som}) can serve as a useful tool to subdivide the high-dimensional colour space into a set of SOM cells efficiently, tracing density and coherent galaxy types, as demonstrated in \citet{Masters_2015} and utilized in \citet{Hildebrandt_2020, Myles_2021} among others. The use of SOMs for redshift calibration demand spectroscopic galaxy samples to completely populate this space, thereby providing accurate galaxy redshifts for any combination of colors. Of critical note is that spectroscopic redshifts are typically obtained only for a specific and limited selection of galaxies and must be weighted to become representative of the photometric sample \citep{Hartley_2020}. The resulting calibration problem is that without a complete spectroscopic sample that fully populates the photometric color-space derived redshift distributions are subject to bias and uncertainty given incomplete or under-sampled spectroscopic observations. 
%Of critical note is that a complete spectroscopic sample fully populating the photometric color-space forms the base of this empirical photo-z calibration scheme, and derived redshift distributions are subject to bias and uncertainty given incomplete or under-sampled spectroscopic observations. 

Some 5,000 spectroscopic redshifts have recently been determined by the Complete Calibration of the Colour-Redshift Relation (C3R2) project in order to populate this color space and thereby calibrate the color-redshift relation \citep{Masters_2017,Masters_2019,c3r2_dr3}. Beyond fully populating each SOM cell, a larger multiplicity of spectroscopic galaxies per SOM cell will be needed to meet requirements for future deep imaging surveys like Euclid \citep{Euclid} and Rubin Observatory \citep{Rubin,lsst_science_v2}. This can be achieved via statistical characterization of broad or bimodal redshift distributions in SOM cells, though a well constructed SOM minimizes these features where possible. Additionally, as upcoming imaging surveys are deeper than most spectroscopic samples, it is essential that the magnitude dependence of the redshift at a fixed color, $dz/dm$, is understood.  Previous examinations of this measurement have shown a small $dz/dm$ trend at fixed color \citep{Masters_2019}. 

In this paper, we present the DESI Complete Calibration of the Color-Redshift Relation (DC3R2) secondary target survey. This survey supplements the existing spectroscopic samples used for photometric redshift calibration by both populating SOM cells that were previously unfilled, and by increasing the multiplicity per cell. We use DC3R2 to revisit the magnitude dependence of the redshift at a fixed color with improved statistics, including apparently brighter galaxies, and study trends in $dz/dm$ as a function of color. 
%but with the higher multiplicity per cell DC3R2 can revisit this with improved statistics, including apparently brighter galaxies, and study trends in $dz/dm$ as a function of color. 
Sec.~\ref{sec:data} introduces the survey data used. In Sec.~\ref{sec:dc3r2data} we describe the construction of the DC3R2 sample. We explore how it calibrates the color-redshift relationship alongside DESI main survey targets in Sec.~\ref{sec:ccrelation}. Finally, using this new resource, we examine the magnitude dependence of redshift at fixed color in the presence of observational effects like photometric scatter (Sec. \ref{sec:magdepend}).

\section{Data}
\label{sec:data}

DC3R2 is a secondary target program on the Dark Energy Spectroscopic Instrument (DESI), Sec. \ref{sec:desi}, that obtained spectroscopic redshifts for galaxies that were targeted with KiDS-VIKING (KV) photometry, Sec. \ref{sec:kids}. \rev{We also analyze DESI main survey targets that fall within the KiDS-VIKING footprint, but were targeted on photometry from the DESI Legacy Imaging Surveys, Sec. \ref{sec:decals}, including the Dark Energy Camera Legacy Survey (DECaLS, \citealt{DESI_Legacy_Survey_DR9}).} Additional spectroscopic and imaging surveys were used to validate this analysis, as listed in Sec. \ref{sec:cosmos}. \rev{For reference, all magnitudes in this paper are absolutely calibrated in the AB system \revtwo{\citep[see][]{AB}}.}

\subsection{DESI}
\label{sec:desi}
\par
The Dark Energy Spectroscopic Instrument (DESI) is a ground-based spectroscopic experiment installed at the 4m Mayall telescope \citep{desicollaboration2016desi_inst, desi_instrument}. The DESI instrument is sensitive from 360-980~nm, with 5,000 robotically actuated fibers that are capable of taking spectra simultaneously. Over five years, it aims to measure spectra of 40 million galaxies and quasars that will aid in examination of baryon acoustic oscillations (BAO), the growth of structure through redshift-space distortions and dark energy \citep{desicollaboration2016desi_science}. While these are DESI's primary goals, the survey is uniquely capable of providing a multitude of spectroscopic redshifts that have far-reaching uses. Here we exploit its ability to be used to calibrate photometric redshifts, in line with the need of weak gravitational lensing experiments.\par
The DESI main survey targets relevant to this paper are divided into three galaxy types: Luminous Red Galaxies \citep[LRGs;][]{LRG_paper}, Emission Line Galaxies \citep[ELGs;][]{ELG_paper}, and the Bright Galaxy Survey \citep[BGS;][]{BGS_paper}. \rev{Notably, LRGs are magnitude selected on a bright \revtwo{z-band} fiber-magnitude alongside $grz$ and NIR photometry from WISE W1 to target red galaxies from $0.4 < z < 1.0$ with high signal-to-noise \cite{LRG_paper}. ELGs, on the other hand, are selected on a $g$-band magnitude alongside $grz$ color cuts to target star forming galaxies from $0.6 < z < 1.6$ \cite{ELG_paper}. Lastly, BGS targets an $r$-band magnitude limited sample (split into a bright and faint subset) that is significantly brighter than the other main surveys, $z \lesssim 0.6$, capable of observation during sub-optimal conditions \cite{BGS_paper}. Each of these main survey samples targets in a complementary part of the color-redshift space, though their union does not span the full census of galaxy populations DESI is capable of measuring. For precise photometric cuts for each sample, we direct the reader to the respective target selection papers, which use the photometry of the DESI Legacy Imaging Surveys \citep{DESI_Legacy_Survey_DR9}. For a brief overview of these selections, see Table \ref{tab:selection_summary}.} 
\par
The methodology for fitting models to the obtained spectra are explored in \citet{redrock}, and the validation of these techniques is performed in \citet{VI_paper}. The observations that produced data for this analysis come from December 14, 2020 through July 9, 2021, which span a combination of the `One-Percent Survey' (OPS), Survey Validation (SV), and the beginning of the main survey operations (Y1) -- the internal\textit{Fuji} and \textit{Guadalupe} data releases, respectively \citep{DESI_SV_Overview}. SV data has already been made available in the DESI Early Data Release (EDR) \citep{desi_edr}. Selection changes for main targets were subject to minute changes between SV1 and the OPS as well as Y1 operations, detailed in the respective paper for each sample. The selection footprint for DC3R2 was modified after the OPS and before Y1 to make use of newly released photometry. Additionally, on May 12th and 13th, 2021 a small dedicated tile program was run for DC3R2 targets with high priority. The selection for DC3R2 targets is outlined in greater detail in Section \ref{sec:targetselection}, and the optimization for dedicated tile fibers is discussed in Appendix \ref{sec:priority}.

\subsection{KiDS-VIKING}
\label{sec:kids}
The Kilo Degree Survey (KiDS) is a large scale optical \textit{ugri} imaging survey with OmegaCAM on the VLT Survey Telescope (VST) at the ESO Paranal Observatory \citep{VST, kids_vst}. Its footprint is overlapped by a near-infrared \textit{ZYJHK$_s$} VIRCAM photometric survey with the 4m Visible and Infrared Survey Telescope for Astronomy (VISTA), the VISTA Kilo-degree Infrared Galaxy Public Survey (VIKING). The two surveys both span more than a thousand square degrees in the \textit{ugriZYJHK$_s$} bands, to a depth of $r\le 25$. Their complementary wavelength coverage has been processed jointly to create the 9-band KiDS-VIKING (KV) survey \citep{Wright_2019}. This data set provides dereddened, multi-band color information for an overlapping patch of the DESI SV and Y1 footprint, allowing us to crucially associate DESI spectroscopic redshifts with a high-dimensional color space. \par 
In this analysis, specifically, we make use of the KiDS-450 data release, with observations spanning the Galaxy And Mass Assembly (GAMA) fields \citep{Driver2011}, depicted as G09, G12, and G15 and the shaded green regions in Fig.~\ref{fig:footprint} \citep{kidsDR3_de_Jong_2017}. Through the beginning of main survey operations, we also used the KiDS-1000 release, shown in the blue footprint of Fig.~\ref{fig:footprint}, \rev{which jointly with KiDS-450 expands the overall footprint and provides a super-set of optical and near-infrared photometry} \citep{Wright_2019,Kannawadi_2019,Hildebrandt_2020}. \rev{The only selections we apply to this calibrated photometry is the provided photometric quality flags (e.g. FLAG\_GAAP\_u) which conservatively mask out (for all nine bands) pixels affected by diffraction spikes, over-saturation, neighboring bright stars, or other effects in the final coadded images. We require all bands to perform our analysis, so any galaxy with flagged poor photometry in a single band is eliminated from our targeting.}

\rev{\subsection{DESI Legacy Imaging Surveys}
\label{sec:decals}
In order to target the main survey samples for DESI, $grz$ optical photometry was collected on across three instruments to produce the DESI Legacy Imaging Surveys  \citep{DESI_Legacy_Survey_DR9}: the Mayall $z$-band Legacy Survey (MzLS) and the Beijing-ARizona Sky Survey (BASS, \citealt{Zou_2017}) at Kitt Peak National Observatory, and the Dark Energy Camera Legacy Survey (DECaLS) at the Cerro Tololo International Observatory. DECaLS extended the photometric catalog of the Dark Energy Survey (DES, e.g. see \citealt{desy3_photom} for photometry or \citealt{des_overview} for an overview), while MzLS and BASS supplied the same filters at complementary high declination ($\delta > +34$). The photometric catalog spans approximately 14,000 deg$^2$ in the northern hemisphere, relevant for DESI and this analysis. Additionally, the optical reach of these surveys was augmented by the Wide-field Infrared Survey Explorer (WISE) bandpasses (3.4 - 22 $\mu$m, W1, W2, W3, W4) \citep{Wright_2010, wise_explan}. More extensive information on the depths and calibration of the Legacy Surveys can be found on the data release page\footnote{https://www.legacysurvey.org/dr9/description/}.}

\subsection{Other Spectroscopic Surveys}
\label{sec:cosmos}

A multitude of spectroscopic surveys have been undertaken in the COSMOS field. Several of these were utilized in this analysis for validation.
%and expand upon past results with the augmented DESI data set. 
Among these surveys are the original C3R2 effort \citep{Masters_2017,Masters_2019,c3r2_dr3} and the master spectroscopic catalog from the COSMOS collaboration (M. Salvato, in prep). The latter  includes observations from a variety of wavelength regimes and spectral resolutions across many instruments (VLT VIMOS, VUDS, Keck MOSFIRE, DEIMOS, Magellan IMACS, Subaru FMOS, and many others, \citep{Lilly2007, LeFevre, Casey_2017,
Hasinger_2018, kriek_2015, kartaltepe_2010, silverman_2015, trump_2007, balogh_2014}. For the use of this project, these samples were limited to only confident redshifts. Furthermore, they were matched to the \cite{Masters_2017} photometry and assigned to the original C3R2 SOM in order to validate our color-redshift relation.
%and compare with DESI spectroscopic redshifts.

\section{Methods and Observations}
\label{sec:dc3r2data}
From December 2020 through July 2021, DESI observed \note{328k} main survey targets (ELGs, LRGs, BGS) in the KV footprint, \note{51,177} of which were also selected as DC3R2 and \note{1216} that were exclusively DC3R2 targets. The following sections will further break down these numbers into the DC3R2 secondary targets during SV ( \ref{sec:targetselection}), Y1 (\ref{sec:sparefiber}), and our dedicated tiles (\ref{sec:dedicatedtiles}), and selected main survey targets that overlap. After completeness cuts we find that we have a sample of \note{230.7k} galaxies that occupy the color space from $0.0 < z < 1.55$. \rev{The aim of this program is to fill as much of the high-dimensional color-magnitude-space for future weak lensing surveys as possible, with high multiplicity and as much depth as is feasible when limited to secondary targets. The sample reported here spans the breadth of 56\% of the population anticipated for Rubin and Euclid analyses, effectively most of the galaxies anticipated at $z < 1.55$. Future work can expand to higher redshift populations on NIR and IR instruments. Crucially DESI is well suited to extending this analysis and examining deeper samples in this lower redshift regime with dedicated programs.}

In the following sections, we describe our procedure for target selection (\ref{sec:targetselection}) through the three major phases of the DC3R2 program: the dedicated tiles (\ref{sec:dedicatedtiles}), and SV, Y1 spare fibers (\ref{sec:sparefiber}). Additionally we detail the observations (\ref{sec:obs}), redshift completeness (\ref{sec:compl}), and the weighting schema to provide a representative sample of redshifts for calibration purposes (\ref{sec:weights}).

\subsection{Target Selection}
\label{sec:targetselection} % used for referring to this section from elsewhere
\rev{
\begin{table*}[]
    \centering
    \begin{tabular}{c|c|c|c|c}
       \hline
       Sample & No. Spec & Redshift Range & Magnitude Cut & Color Selection \tabularnewline
       \hline
       DC3R2  & 51,177 & $0 < z < 1.6 $& $z_{\mathrm{fiber}} < 22.10$ & SOM, KV $ugriZYJHK_s$\\
       LRG  & 46,771 & $0.4 < z \lesssim 1.0 $ & $z_{\mathrm{fiber}} < 21.60 $  & $grz$ + $W1$ \\
       ELG  & 125,027 & $0.6 < z < 1.6 $& $g > 20$, $g_{\mathrm{fiber}} < 24.1$ & $grz$ \\
       BGS bright & \multirow{2}{*}{\centering 156,895} & $0 < z \lesssim 0.6 $& $r < 19.5$, $r_{\mathrm{fiber}} \lesssim 22.9$ & magnitude limited\\
       BGS faint &  & $0 < z \lesssim 0.6 $& $19.5 < r < 20.175$, $r_{\mathrm{fiber}}\lesssim 22.9$& $grz$, + $W1$\\
       \hline
    \end{tabular}
    \caption{\rev{Generalized target selections for each survey contributing to this analysis and their complementary redshift sensitivities and the resulting observed spectra counts before quality cuts are applied. All magnitude cuts are reported in the \revtwo{optical Legacy Survey bands, $grz$, made use of for targeting}. Note that the BGS fiber magnitude cuts are color dependent and simplified here, and that objects may be shared between different target classes. See Fig. \ref{fig:magbins_all} for a breakdown of the interplay of each sample in color-magnitude space.}}
    \label{tab:selection_summary}
\end{table*}
}
The primary DC3R2 targets span across the GAMA-9h, 12h, and 15h equatorial fields \citep{Driver2011} for a total of 300 sq. degrees, where sufficiently deep $ugriZYJHK_s$  color information was available. We have matched the GAMA fields as reported by KiDS-VIKING (KV) DR3 \citep{Wright_2019} to the Dark Energy Camera Legacy Survey (DECaLS; \citealt{DESI_Legacy_Survey_DR9}) Data Release 9 photometry in order to constrain color alongside DESI \revtwo{z-band} fiber fluxes, using the closest match within 1". \rev{The fiber fluxes from DECaLS were used to ensure our program could be completed at low exposure cost.} In order to calibrate the complete color redshift relation, DC3R2 aimed to occupy with multiplicity as much as reasonably attainable to DESI of the color-space described by the \citealt{Masters_2017} Self Organizing Map (SOM), which is a useful map trained on narrow band COSMOS photometry that spans the approximate depth and breadth of color for future weak lensing surveys like Euclid and LSST. Appropriately it is the subject of several redshift search programs that also aim to span this color-space with spectroscopy \citep{euclidprep,Masters_2015}, which further encourages our use of it. This SOM is then corrected to KiDS-VIKING colors according the procedure described in Section \ref{sec:verification}, to allow for assignment of our alternate photometric bands. The abundances of galaxies observed across the color-space for these KiDS-450 fields (G09, G12, G15) can be found in Fig. \ref{subfig:abundance}. \rev{In the neighboring panel, Fig. \ref{subfig:succ_spec} depicts the distribution of galaxies observed from this catalog with successfully measured redshifts.} 

Galaxies are selected by color, determined by their assignment to cells in this map, as well as by fiber magnitude. We select these colors on mean cell magnitude, to take into account visibility of the targets to the fiber, and on a high probability of redshift < 1.6 (i.e. the \rev{[OII]} feature lies within the DESI wavelength window\rev{, discussed further in the following paragraphs}). This allows us to target 3,692 cells from the C3R2 SOM ($\approx$33\%) with a minimum of 3 galaxies per cell, spread in magnitude. These targets were chosen to enable a first quantification and rejection of faulty photometry and redshift and examine the trend of redshift with magnitude at fixed color, sorely required for accurate calibration efforts in the future \citep{Masters_2019}. From 1800 targets per sq. deg. DC3R2 required only a small random subset (36 per sq. deg., or 2\% for our initial request of 3 galaxies per cell, making it an ideal candidate for a secondary target programs with very flexible fiber assignment across a high density of targets. \rev{While ideally, DC3R2 would sample the full color space available to DESI, i.e. no selection on $z < 1.6$, to fully probe the edge of the redshift range without a photometric redshift prior, these galaxies are realistically too faint on average for a spare fiber program to address in a few 20 minute exposures. We benefit from main survey target classes (ELGs especially) probing this redshift regime with higher fidelity. We can summarize the diverse range of target selections in this paper in Table \ref{tab:selection_summary}, where we find DC3R2 is the broadest in terms of redshift, but augmented by deeper main survey samples.}\par

Explicit choices for the DC3R2 target selection procedure follows from the joint KiDS-VIKING-DECaLS matched catalog,
\begin{itemize}
    \item To boost redshift completeness of our targets, individual galaxies that would conservatively be expected to take four visits or fewer were selected from the catalog as MAG\_FIBER\_Z < 22.10 \rev{-- too acheive a balmer break SNR of approximately one in the optical, where passive galaxies are the conservative upper bound, see the discussion in \citealt{LRG_paper} (their Sec. 4.4) for more information.}.
    \item Potential targets were assigned to the \citet{Masters_2017} SOM, after transforming the latter to the KiDS-VIKING $ugriZYJHK_s$ color-space (see Sec.~\ref{sec:verification}).
    \item An envelope of SOM cells (not contiguous) that contain galaxies bright enough to achieve redshift success with two visits  (mean(MAG\_FIBER\_Z) < 21.88 and >95\% probability of redshift < 1.6) were selected. The probability of redshift < 1.6 for a given cell was obtained from the COSMOS15 \citep{Laigle2016} photometric redshifts for each cell as in \citet{Masters_2015}.
    \item The full target catalog was defined as the concatenation of all cells within the above envelope that each had at minimum 3 viable detections (believed to have observable redshifts in four or fewer DESI visits). From this we prioritized the brightest in each cell (as a lever arm for the measurement in Sec. \ref{sec:magdepend}) and multiplicity as described for each component of our survey in the respective section below.
\end{itemize}
The initial spare fiber catalog for the SV stage had approximately 970k targets available that met the above criteria. This target list was modified for Y1 to include the larger KiDS-1000 footprint \citep{KiDS_DR4}, see~\ref{sec:sparefiber} for more information. \rev{Our ability to prioritize targets is severely limited by the secondary nature of our program, and where we do have ability to prioritize is discussed in more depth in App. \ref{sec:priority} for the dedicated tiles. We are agnostic to pre-existing spectroscopic coverage in the field and simply cover any available target by chance target-fiber vacancies.}

\begin{comment} %outdated figure
\begin{figure}
    \centering
	\includegraphics[width=0.45\columnwidth]{draft_figures/potential_coverage.png}
    \caption{The targeting abundance for objects with KV colors that are visible to DESI. \note{swap to three figure SOM with abundance, successful spec, and bad spec}}
    \label{fig:abund}
\end{figure}
\end{comment}

\subsubsection{Dedicated Tiles}
\label{sec:dedicatedtiles}

DC3R2 has observed two dedicated tiles, i.e. instrument pointings, at the end of the One Percent Survey within the survey validation period. We chose the two pointings to be centered at 217.5 degrees and 221.0 degrees RA on the celestial equator, within the GAMA 15h field \citep{Driver2011} and also within the Hyper Suprime-Cam Subaru Strategic Program (HSC) survey area \citep{Aihara2018}. 
Each tile was observed with two different fiber configurations, one for a single 30 minute exposure aiming at targets down to \revtwo{$z_{\rm fiber} < 21.5$}, and one for a total of 90 minutes of exposure time with targets down to \revtwo{$z_{\rm fiber} < 22.10$}. A description of how targets were chosen for each of these pointings can be found in Appendix \ref{sec:priority}, \rev{but in summary we optimize priority for bright-faint pairs (a large span of MAG\_GAAP\_Z) in a given cell to maximize our lever arm on measuring $dz/dm$}. The dedicated tiles are identified in Fig.~\ref{fig:footprint} as black circles.

\subsubsection{Spare Fibers}
\label{sec:sparefiber}

The DC3R2 targets observed outside of the dedicated tiles occurred in two phases within the SV and Y1 periods. During SV, DC3R2 took observations for \note{44,272} targets. In contrast, during the onset of Y1 \note{6,905} DC3R2 objects were observed.

The initial spare fiber, \textit{secondary target}, observations are described by the selections made in \ref{sec:targetselection} on the KV DR3, KiDS-450, photometry released at the time.

The spare fiber target selection and prioritization for the Y1 observing period was modified from the initial SV phase to draw targets from the then publicly available larger area KiDS-1000 (DR4) photometric catalog \citep{KiDS_DR4}, as seen in the shaded blue regions of Fig. \ref{fig:footprint}.
\begin{comment}
In light of having a dedicated tile run that ought to give an appropriate depth, the catalog of targets for the DC3R2 survey was adjusted from the strategy described in Section \ref{sec:targetselection}, during SV, for DESI Y1 in order to obtain a wider sample of bright galaxies over a larger footprint. The KiDS-1000 (DR4) photometry \citep{KiDS_DR4} was made available at this time, as seen in the shaded regions of Fig. \ref{fig:footprint}, and our target footprint was expanded for Y1 accordingly.
\end{comment}
The strategy was altered to prioritize the brightest galaxy in each cell alongside a randomly drawn target within a Z-magnitude of 21.88. This lower magnitude cut ensures a redshift is obtainable from two visits, which differs from the maximum of four visits in SV, seldom available for spare targets. The random draw traces the magnitude distribution of the cells overall. This pair-wise selection per cell has the benefit of extending the magnitude leverage on $dz/dm$, as more bright galaxies are available in the larger footprint, and achieving high redshift success in survey spare fiber mode.
\begin{comment}Since providing the brightest galaxy per cell is highly desirable to obtain large magnitude separations in a SOM cell, and therefore constrain $dz/dm$, it is preferable to have a wider survey area for bright galaxies than the dedicated tiles. Additionally as brighter objects are less common, a larger footprint is better poised to identify them than two pointings. 
\end{comment}
The same criterion as in Sec.~\ref{sec:targetselection} allowed us to select from 5074 cells in the C3R2 SOM for the Y1 fiber proposal. The blue objects in Fig. \ref{fig:footprint} are successful matches back to this wider catalog during Y1, sometimes retroactively from main survey targets in the SV phase. The DC3R2 spare fiber program extended into Y1, and while some of that data is analyzed here, we expect an additional \note{315k} targets within our footprint to become available with the Y1 data release, with around \note{28k} of those objects being DC3R2 exclusive targets and the remainder coming from overlap with DESI main survey target classes.

\subsection{Observations}
\label{sec:obs}
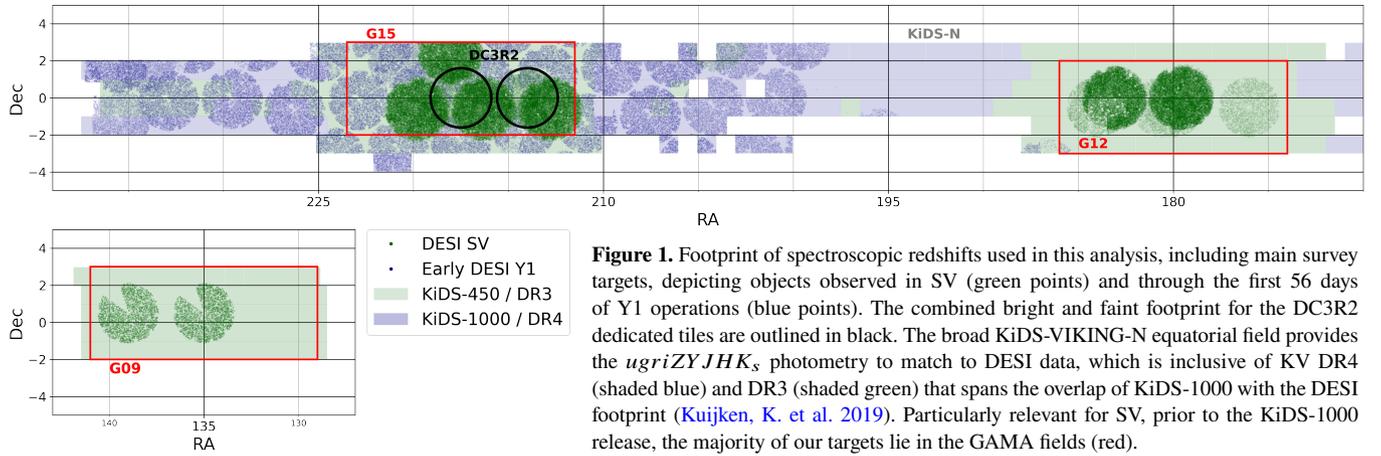
\begin{figure*}
    \centering
	\includegraphics[width=\textwidth]{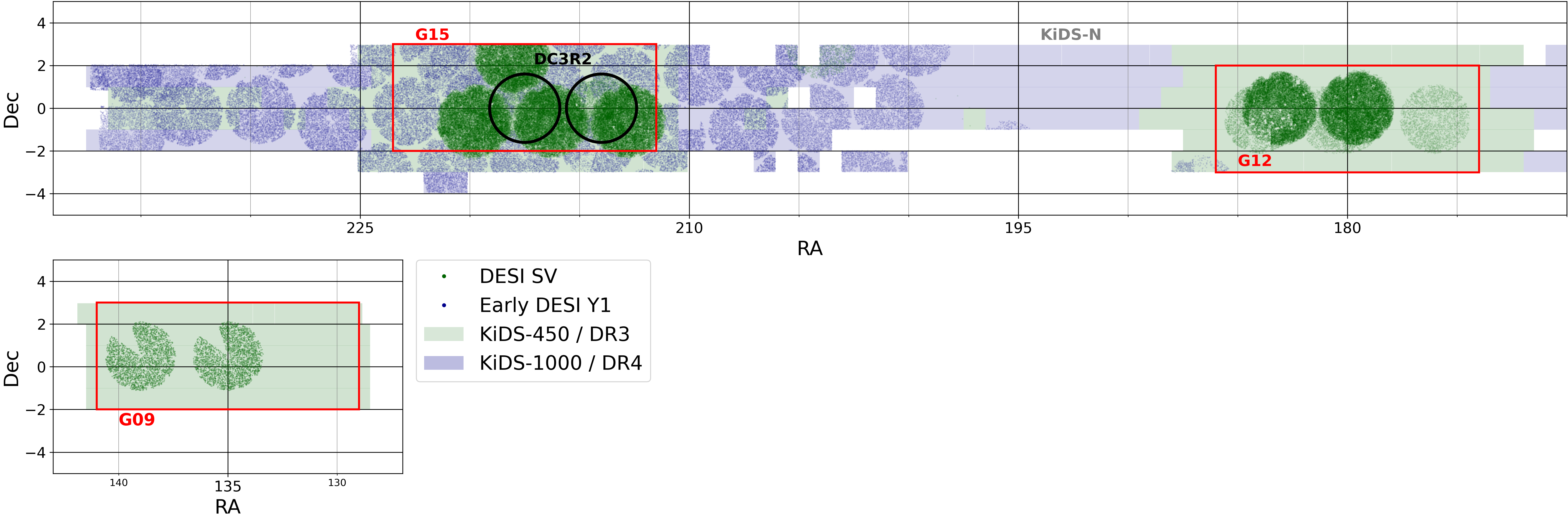}%
    \begin{picture}(0,0)
    \put(-289,37){\parbox[h]{0.565\linewidth}{\centering
    \caption{Footprint of spectroscopic redshifts used in this analysis, including main survey targets, depicting objects observed in SV (green points) and through the first 56 days of Y1 operations (blue points). The combined bright and faint footprint for the DC3R2 dedicated tiles are outlined in black. The broad KiDS-VIKING-N \rev{equatorial} field provides the $ugriZYJHK_s$ photometry to match to DESI data, which is inclusive of KV DR4 (shaded blue) and DR3 (shaded green) \rev{that spans the overlap of KiDS-1000 with the DESI footprint} \citep{KiDS_DR4}. Particularly relevant for SV, prior to the KiDS-1000 release, the majority of our targets lie in the GAMA fields (red).}
    \label{fig:footprint}}}
    \end{picture}
\end{figure*}

The positions of all observed DC3R2 targets on the sky, \rev{falling on the equatorial where 9-band photometry is available,} are depicted in Fig. \ref{fig:footprint}, including the dedicated tiles in black. \rev{An individual DESI pointing can capture up to 5k targets across 8 deg$^2$, which is visible as the overlapping circular rosettes in this figure.}

\begin{figure*}
    \centering
    \begin{subfigure}[t]{0.285\textwidth}
        \includegraphics[width=\textwidth]{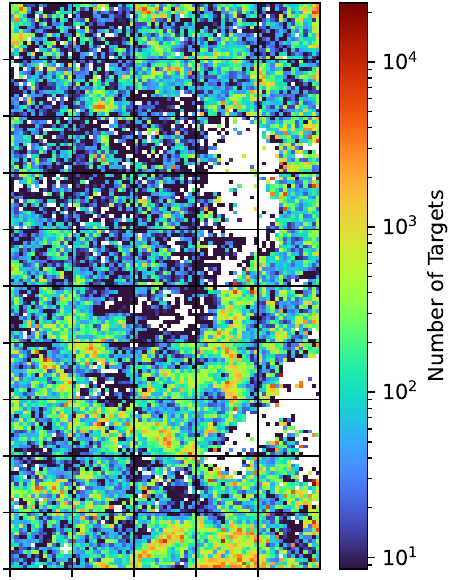}
        \caption{Target abundance by $ugriZYJHK_s$ color self-organizing map cell in the relevant KiDS-VIKING 450 fields (G09, G12, G15), from which spectroscopic targets for DC3R2 were chosen for SV and the dedicated tiles. Objects included in this count meet either DC3R2 selections or any main survey (ELG, LRG, BGS) magnitude cuts in \textit{grz}.}
        \label{subfig:abundance}
        \end{subfigure}
    \hspace{0.01\textwidth}
    \begin{subfigure}[t]{0.285\textwidth}
        \includegraphics[width=\textwidth]{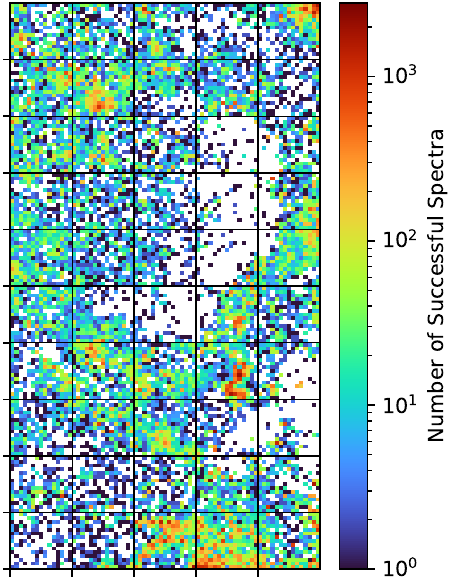}
        \caption{Number of successful spectra taken per cell, across DC3R2 as well as all DESI main surveys (ELG/LRG/BGS) that pass redshift quality cuts.}
        \label{subfig:succ_spec}
        \end{subfigure}
    \hspace{0.01\textwidth}
    \begin{subfigure}[t]{0.285\textwidth}
        \includegraphics[width=\textwidth]{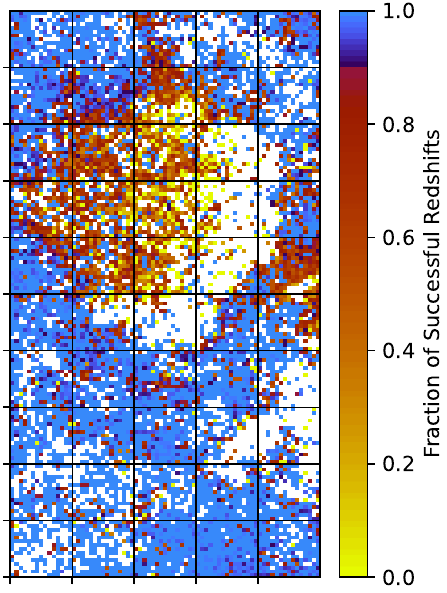}
        \caption{Fraction of spectra that meet our redshift quality selections at a given color, across all surveys, out of all objects observed. If the fraction of spectra in the cell with certain redshifts amounts to less than 90\% we consider the spectra that are measured incomplete and potentially biased and exclude the cell from the fiducial analysis.}
        \label{subfig:zcomplete}
    \end{subfigure}
    \caption[t]{The distributions of targeting, spectroscopy, and redshift completeness across the color-space.}
    \label{fig:threesom}
\end{figure*}

\begin{figure*}
    \centering
    \begin{subfigure}[t]{0.285\textwidth}
        \includegraphics[width=\textwidth]{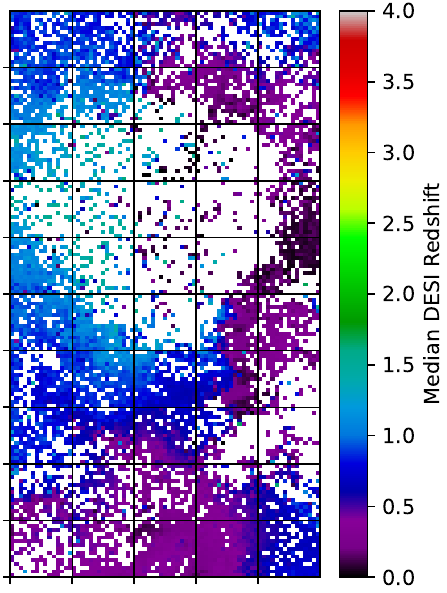}
        \caption{Median spectroscopic redshifts for a given cell for all DC3R2 and main survey targets that meet our quality cuts.}
        \label{subfig:zmedians}
        \end{subfigure}
    \hspace{0.01\textwidth}
    \begin{subfigure}[t]{0.285\textwidth}
        \includegraphics[width=\textwidth]{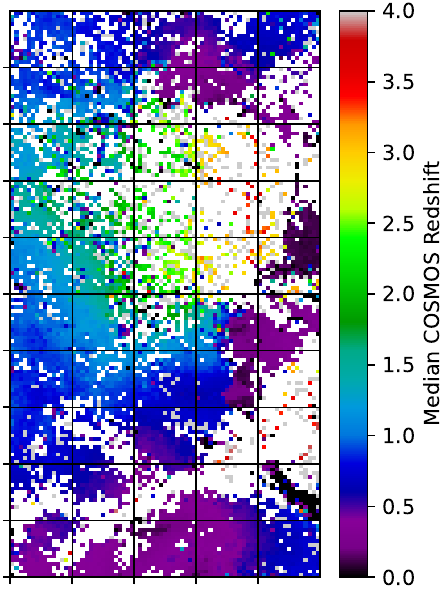}
        \caption{Median spectroscopic redshifts for a given cell for the COSMOS field (as discussed in \ref{sec:cosmos}), from prior, non-DESI spectroscopy.}
        \label{subfig:zmedians_cosmos}
        \end{subfigure}
    \hspace{0.01\textwidth}
    \begin{subfigure}[t]{0.341\textwidth}
        \includegraphics[width=\textwidth]{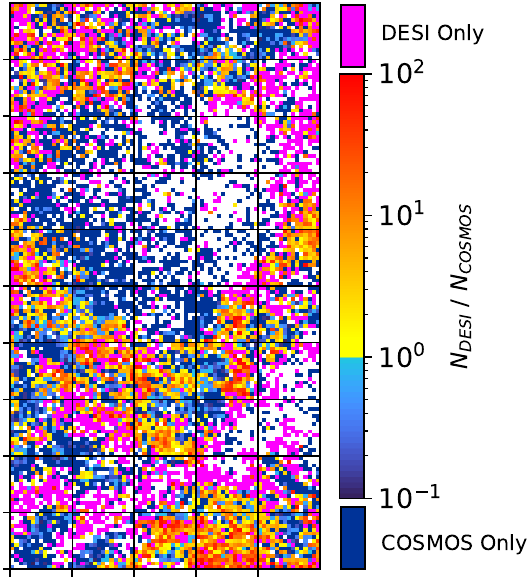}
        \caption{The ratio of spectroscopic redshift counts between initial DESI efforts and prior COSMOS data for the same map, where solid red depicts color space exclusively added by DESI and dark blue does the same for COSMOS. The transition occurs where DESI doubled existing spec-\textit{z} counts.}
        \label{subfig:countdiff}
    \end{subfigure}
    \caption[t]{Median DESI redshifts compared to previously measured and curated spectroscopy in the COSMOS field reveals substantial gains in the low redshift regions and overall multiplicity. Note that these color-spaces are not identical and this comparison is a generalized one, as per the discussions in Sec. \ref{sec:verification}, Appendix \ref{app:som_verif}.}
    \label{fig:zmedians}
\end{figure*}

\subsubsection{Redshift determination}

Following pixel level calibration and extraction of a one-dimensional spectrum, the DESI redshifting pipeline \textsc{redrock} forward models the observed data from a basis consisting of a set of template SEDs for each target class \citep{desi_spec_pipeline,redrock}. This decomposition is done at each point in a fine grid of redshift values $z$ and the $\chi^2$ of the difference of observed data and the best linear combination of templates is determined. A minimum in the $\chi^2(z)$ indicates a potentially optimal redshift-template solution. The $z$ that globally minimizes $\chi^2$ is taken as the redshift of the observed object.

\subsubsection{Completeness}
\label{sec:compl}

The primary metric of redshift confidence in the DESI survey is $\Delta \chi^2$, the difference between the $\chi^2$ of the best fit redshift and template combination and the second lowest local minimum of $\chi^2$ as a function of redshift. The larger the $\Delta \chi^2$, the more confident one can be that the first redshift is correct \citep{redrock}. As each main survey sample selection (\citealt{BGS_paper}, \citealt{LRG_paper}, and \citealt{ELG_paper}) has different characteristic SED features, the ability of \textsc{redrock} to fit a given spectrum depends on the type of galaxy. The DESI visual inspection efforts have determined minimum $\Delta\chi^2$ selections and additional quality cuts that maximize purity and completeness for each main survey sample \citep{VI_paper, desi_spec_pipeline}.

For this analysis, DC3R2 galaxies that we consider to have high confidence redshifts must pass the same completeness criteria as the BGS sample, i.e. $\Delta \chi^2 > 40$. This is the strictest among the $\Delta \chi^2$ cuts for DESI main survey samples and was chosen to account for the broad variation of SED types in our target selection, \rev{informed by the visual inspection done for the BGS sample in \cite{VI_paper} that optimizes good redshift purity and inclusion for our comparably bright targets}.  ELGs and LRGs have significantly lower $\Delta \chi^2$ requirements in the DESI main survey, which combined with further quality cuts give high fractions of confident redshifts, as \textsc{redrock} and its template sets have been optimized to obtain good fits to these SEDs. Galaxies with different SEDs may need more conservative metrics to minimize outliers, hence our adoption of the $\Delta\chi^2>40$ cut. \rev{Our analysis indicates this is more than sufficient in most regions of color-space, but it ought to be kept in mind that $\Delta\chi^2$ is only one avenue of selecting robust redshifts and the threshold required depends strongly on the SED type of the galaxy. Other features, e.g. the signal-to-noise in strong features like [OII], may also be strong predictors of successful redshift measurements in \textsc{redrock}.}
\par
\rev{We report the distribution of successful redshifts in the SOM from all target classes based upon their individual success criteria in Fig. \ref{subfig:succ_spec}. While a cell may have successful redshift measurements, those measurements may be biased in certain regions of color-space where obtaining good redshifts is more difficult. How we account for this bias is described in the following section. The redshifts measured from these successful spectra are visible in Fig. \ref{subfig:zmedians} after accounting for these incomplete regions. We can see in Fig. \ref{subfig:zmedians_cosmos} the general agreement with previous spectroscopy in the COSMOS field. The color-redshift calibration is further discussed in Sec. \ref{sec:colorspace}.}

\begin{figure}
    \centering
	\includegraphics[width=0.7\columnwidth]{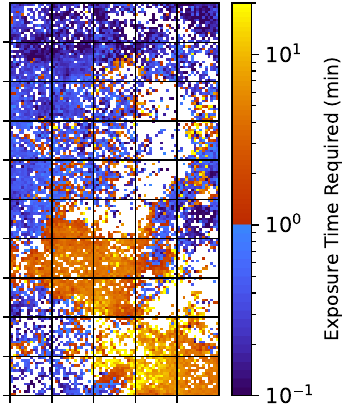}
    \caption{Median DESI exposure times (in minutes) necessary for each SOM cell, for targets with a DECam fiber magnitude of $z_{\rm fiber}=21$, to reach a high redshift confidence $\Delta \chi^2 = 40$, as per Appendix \ref{sec:exptime}. Full exposure time quantiles and median colors for each cell are reported in the attached data products. BGS targets are excluded from this plot, due to their typically much larger sky brightness. \rev{The color bar is chosen to approximately separate quiescent, passive galaxies (orange)} from emission line galaxies (blue).}

    \label{fig:exptime}
\end{figure}

For the sake of future survey efforts to calibrate the color-redshift relation, in Fig. \ref{fig:exptime} we report the exposure times required by DESI for targets of eight-band colors in the SOM to achieve $\Delta \chi^2 > 40$ for objects scaled to a magnitude of $\mathrm{MAG\_GAAP\_Z} = 21.0$, as following from the discussion in Appendix \ref{sec:exptime}. \par
For the fiber configurations with 90-minute exposure time over the DC3R2 dedicated tiles, 98.1\% of targets achieved the DC3R2 criterion for success with no flagged warnings ($\mathrm{ZWARN == 0}$). For the galaxies in the two single 30-minute exposures, \note{93.4\%} meet this redshift success criterion. The overall redshift success rate for the dedicated tile targets is \note{95.8\%}. For the entire DC3R2 ancillary program that also made use of spare fibers (footprint depicted in Fig. \ref{fig:footprint}), the success rate is \note{93.4\%} for unique DC3R2 targets, and \note{94.7\%} for shared targets, with a total of \note{13,270} targets observed during the dedicated program. While many of these overlap with main survey targets, the higher DC3R2 priority in fiber selection ensures that the selection of these objects are less biased towards the main survey selections in color than similar overlaps in the spare fiber fields.\par

\begin{comment}
\subsubsection{Examination of Outliers}
\label{sec:outliers}
\DG{needs to be more descriptive of what exactly has been done}

Contamination of the spectroscopic sample are potential concern for redshift calibration. Whether they be from misidentified objects like stars or AGN, or poor model fits to observations, looking at outliers on the both the level of individual spectroscopy as well as aggregate issues in color space is a useful tool to identify a pure sample. \note{Second best redrock fits for poor cells with poor $\sigma_z$, plot of best fit spectra in SOM cells that are bimodal, or have dramatic outliers/high stdev.} 
\end{comment}

\subsubsection{Sample Weighting Scheme}
\label{sec:weights}
One of the primary methods of photometric redshift calibration in weak lensing is to appropriately reweight a sample of known redshifts to accurately approximate the redshift distribution of a much larger source galaxy sample. This requires a full understanding of the selection acting on both samples. For the true redshift sample this selection can be intentional in the definition of a spectroscopic target catalog or unintentional due to incomplete redshift recovery among the targeted galaxies. Only with the selection accounted for an in addition with the weak lensing source galaxy sample selection applied can the reweighted spectral sample be representative and the estimated redshift distribution therefore be unbiased.

For this survey, we must take into account the overlap of DC3R2 targets with DESI main survey targets. The selection of the latter is not based on KiDS-VIKING colors but on DESI Legacy Survey $g,r,z,W1,W2$ photometry \citep{targsel_paper}. The observation prioritization of the DESI main survey targets may affect relative abundances of certain redshifts within a given SOM cell. This is accounted for by reweighting according to different subselections of our targets, namely:
\begin{enumerate}[label=(\arabic*),leftmargin=2\parindent]
    \item \textbf{Dedicated tiles}: Our dedicated tile sampling method over a small area will be unaffected by DESI main survey oversampling. It is our \textit{fiducial} sample for this reason.
    \item \textbf{DC3R2 exclusive targets}: These are targets observed during regular DESI operations that are only observed as targets of our dedicated program, i.e.~that do not meet DESI main survey target selection criteria. This sample will be potentially biased \emph{away} from the types and redshifts of galaxies observed by DESI's main surveys in SOM cells that contain a mix of both.
    \item \textbf{DESI main survey targets}: DESI main survey targets that overlap with our selection will be more likely to be observed due to their high priority in comparison with DC3R2 exclusive targets in SOM cells that contain a mix of both. Each class of main survey targets has different magnitude and color cuts than the DC3R2 targets. SV and Y1 main survey targets have different color selections that have evolved over the course of the survey and are weighted accordingly.
\end{enumerate}
We can re-weight these subselections to leverage the full DESI survey beyond our fiducial sample, (1). The basic ruleset for usage and weighting is described here, with the aim to be providing a reliable and unbiased sample of spectroscopic galaxies over as many cells as possible with a large magnitude span within available cells. We want each cell to have a collection of representative galaxies that are corrected for overabundances caused by preferential observations of galaxies of a particular SED type. For all DESI-observed galaxies that have KiDS-VIKING photometry in a given cell:
\begin{itemize}
    \item We split the sample in the cell into categories of (2) and (3) by observation flags $\mathrm{DESI\_TARGET}$ and $\mathrm{SCND\_TARGET}$, \rev{available bitmasks in the DESI catalog that indicate main survey and secondary programs respectively}. 
    \item For those in (2), we apply a cut on redshift confidence and warning flags ($\mathrm{DELTACHI2} > 40$, $\mathrm{ZWARN}=0$).
    \item For those in (3), we check how complete the \textit{targeting} is for main survey targets in color space by comparing the full main survey target catalog occupation for the cell (with only color cuts applied) to the full KiDS-VIKING occupation of that cell. If the targeting completeness (fraction of photometric targets that are also main survey targets) is below 90 per-cent for a given color, we exclude the main targets from this cell and skip the following steps. This effectively ensures that the color cut on the main survey sample does not significantly bisect a SOM cell. As the period covered by the observations spanned several iterations of LRG/ELG cuts (see \citealt{targsel_paper}), only cells where both the stricter Y1 cut and the looser SV cuts did not bisect the cell were retained.
    \item For those in (1), (2) and (3), we check the redshift completeness of the sample type in the given cell. This is done on the subset of galaxies in a cell that pass the specific survey target selection (e.g. the DECaLS based magnitude and color cuts, or DC3R2 magnitude selection). If among this subset within a cell the redshift completeness is less than 90\%, we exclude all of these for both DC3R2 and main survey galaxies from the weighting scheme due to the risk of a significant redshift dependent selection bias, and do not include low confidence targets in the next step. See a simple visual description of this in Fig. \ref{subfig:zcomplete}. \rev{The large number of incomplete spectra seen in the center top left of this figure do not lack sufficient exposure time, but spectral features within the DESI sensitivity range. The galaxies likely to be $z > 1.6$ predominantly lack spectral features in the optical.}
    \item We weight the samples of (2) and (3) high confidence redshifts so that both samples contribute proportionally to their abundances in the targeting catalog, within a given cell and bin of $z_{\rm fiber}$. If only one sample type exists in a given magnitude/cell bin, no such re-weighting is done.
\end{itemize}

\begin{figure*}
    \centering
    \includegraphics[width=0.7\textwidth]{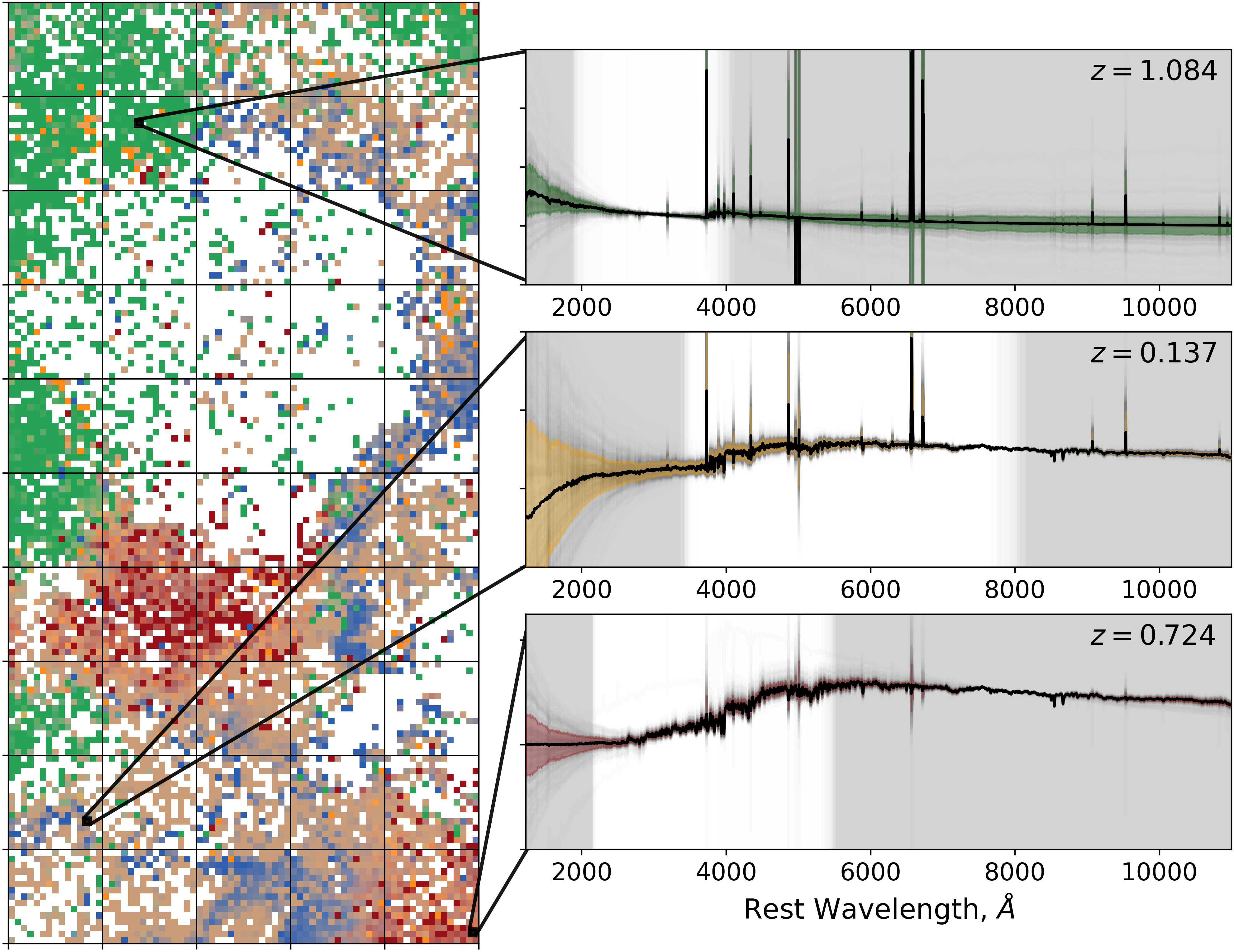}
    \caption{Depiction of the SOM with four complementary samples, post completeness cuts:  Emission Line Galaxies (green), Luminous Red Galaxies (red), the Bright Galaxy Survey (blue) and DC3R2 selected targets (orange), where the strength of each color channel is directly proportional to the fraction of galaxies in that cell that come from the sample noted. The distributions of normalized \textsc{redrock} SED fits are shown for three choices of SOM cell \rev{that span star forming, star burst, and passive galaxy types}, with a colored envelope for the 68\% quantile region about the median of all templates shown in black. The shaded regions denote rest-frame wavelengths not observed by DESI according to the best fit redshift \rev{and individual galaxy spectrum fits are drawn in gray}. Normalization is done \rev{at a point of featureless continuum}, 3000 Angstroms for star forming galaxies and at 8000 Angstroms for mostly quiescent galaxies.}
    \label{fig:joint_coverage}
\end{figure*}

We see a linear relationship be written explicitly as follows. For a given magnitude/color bin $b$ within the $g,r,z$ magnitude cube of a given SOM cell $c$ let $N^{a}$ denote the number of actual (or \textit{observed}) and $N^{p}$ the number of potential targets. Further, denote the unique sample (ELG, LRG, BGS (B), BGS (F), DC3R2 spare fibers, DC3R2 dedicated tiles) of a target by the index $s$. Where each class of galaxy lives in the SOM is delineated by color in Fig. \ref{fig:joint_coverage}, where we can see each of the samples are highly complimentary. The weight of the redshift of an object towards the estimated redshift distribution of a cell $c$ depends on $s$ and $b$ and can be factorized as
\begin{equation}
    \centering
    w(b,c,s) = w_{b}(b,c)\times w_{s}(b,c,s)
    \label{eq:weights}
\end{equation}
where $w_{b}(b,c)$ is the weight of the magnitude bin towards the full sample in the cell and $w_{s}(b,c,s)$ is the weight of the sample that the galaxy belongs to for the given cell and magnitude bin. Individual samples in magnitude-color bins require these weights to properly reproduce relevant target abundances in our final reported redshift distributions. An illustration of how samples populate a choice of magnitude bins is visible for the SV3 selections in Appendix \ref{app:som}, Fig. \ref{fig:magbins_all} where a more thorough description of magnitude bins can be found. For objects that are in more than one sample (typically only true for some DC3R2 targets with main survey samples), the largest of its sample weights is used.
\begin{equation}
    w_{b}(b,c) = \frac{N_b^p / N_c^p}{N_b^a / N_c^a}
    \label{eq:wb}
\end{equation}
Here $N_{c}^{a,p}$ and $N_{b}^{a,p}$ refer to the number of actual or potential targets in cell $c$ or in magnitude bin $b$ of that cell, respectively. Similarly, the reweighting of samples among that bin and cell is given by

\begin{equation} \label{eq:ws}
    w_{s}(b,c,s) = \frac{N_{s,b}^a / N_b^a}{N_{s,b}(\rm{spec}) / N_b(\rm{spec})}
\end{equation}

\begin{figure}
    \centering
    \includegraphics[width=\columnwidth]{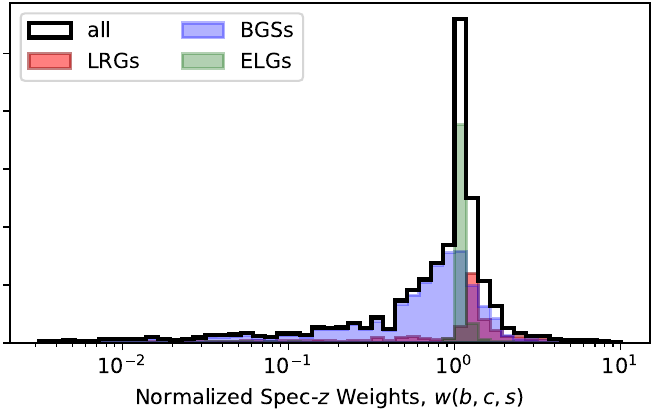}
    \caption{Distribution of the weights for main target classes. Over-represented spectroscopic targets will have lower weight by construction, in order to generate a more representative $p(z)$.}
    \label{fig:weight_hist}
\end{figure}

No reweighting is done for objects in (1) aside from the redshift completeness check, as these targets are not expected to be biased within a cell. The distribution of weights normalized for each SOM cell, $\Sigma_{s\in c} \Sigma_{b\in c} w_s(b,c,s) = N_{c}(spec)$, are depicted in Fig. \ref{fig:weight_hist} for the main survey samples. The purpose of these weights, to construct less biased $p(z|c)$ distributions, can be demonstrated in Sec. \ref{sec:bias}, where they produce noticeable shifts in the inferred redshift distributions.

\begin{comment}
{\DG{I'm confused now - I would have thought $t_t$ and $spec$ indicate the same thing, actually observed targets, and would have expected to see a "potential over actual" ratio like in eqn 4 again, but now among samples within the bin. What am I missing?}
\begin{figure}
    \centering
    \includegraphics[width=\columnwidth]{draft_figures/color-blocks.PNG}
    \caption{Magnitude bins for the reweighting scheme are depicted above for a single slice in Z-band magnitude, with a description of which samples are present in each color-bin.}
    \label{fig:magbins}
\end{figure}
\end{comment}

\section{The Color-Redshift Relation}

In this section we describe the color-space used in this analysis (Sec. \ref{sec:verification}) and discuss the characterization of redshift distributions in that space given DESI observations (Sec. \ref{sec:colorspace}).

\label{sec:ccrelation}
\subsection{Description of Color-Space}
\label{sec:verification}

This analysis makes use of the SOM first developed in \cite{Masters_2015}, which includes galaxies to Rubin and Euclid-like depth, $i\lesssim24.5$ or approximately 98\% complete at $i=25.3$ from \citealt{Laigle2016}, in  \textit{ugrizYJH}, with the modifications and added $K_s$-band from \cite{Masters_2017}. \rev{Our map exists in analogous 9-band KiDS-VIKING photometric colors to the original training photometry. As at the time of this paper KiDS-VIKING-like photometry is not available across the same fields, our transfer function to shift the map is evaluated by measuring flux in the KiDS-VIKING bandpasses for the best model template fit to the narrow band photometry in the in the original SOM cell. The nature of the template fitting and the templates themselves to produce these SEDs are elaborated upon in \citet{Laigle2016}. While KV is} not the photometry that \citealt{Masters_2015} was trained on, though making use of a transformed version of this SOM has the advantage that individual cells will be populated by galaxy populations of approximately the same true colors. Thus the deep samples collected by C3R2 and the DC3R2 survey will have similar redshifts for similar cells and can help inform future redshift searches. We can conclude that our map, though photometrically different, will span a similar color space to \citealt{Masters_2017} and therefore Euclid and LSST by construction. For more elaboration on the differences between the SOM in this analysis and \citealt{Masters_2017}, see Appendix \ref{app:som_verif}, and for the exact colors of our map see Appendix \ref{app:som}.

\subsection{Characterization of Color-Space Using DESI}
\label{sec:colorspace}
With DESI targets mapped to $ugriZYJHK_s$ color-space, we have a powerful statistical sample that constrains the high-dimensional color-redshift relation for future cosmological analyses. With more than \note{230k galaxies covering 56\%} of the map, we present redshifts that can span 0 < \textit{z} < 1.6 with high multiplicity. DESI provides both secure spectroscopic redshifts alongside \textsc{redrock} SED models fitted to each galaxy, which allows us to explore the evolution of galaxy-type across the map, as seen in Fig. \ref{fig:joint_coverage}. The colors of the SOM, broadly have smooth transitions, as do the redshifts as seen in Fig. \ref{subfig:zmedians} which depicts the median redshift across this map. Galaxies with alike colors tend to have alike redshifts, except for where sharp delineating features express an innate degeneracy - where small shifts in color can have large consequences for redshift inference. \par
\begin{figure}
    \centering
    \includegraphics[width = \columnwidth]{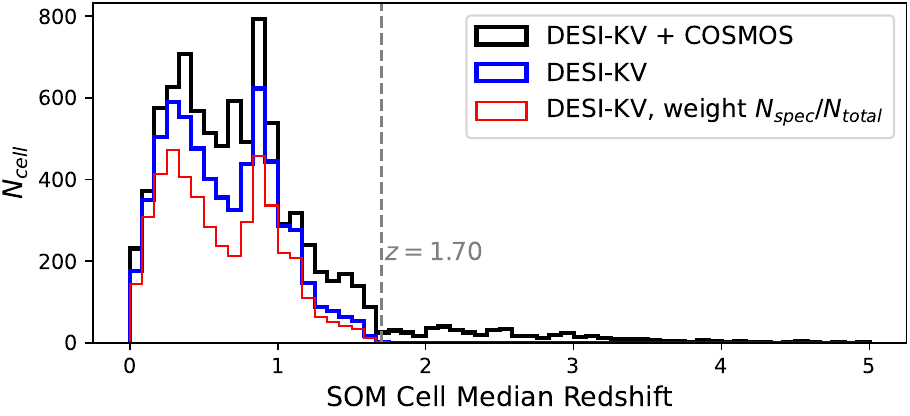}
    \caption{Existing spectroscopic coverage of the Rubin/Euclid color space, shown as a histogram of SOM cell counts against their median redshifts. Contributions from this work (DESI + KV photometry, blue), are combined with data taken in the COSMOS field for reference (black). The contribution of DC3R2 populated cells, weighted by the count of DESI spectra per combined spectroscopic count (red), demonstrate that jointly DC3R2 alongside DESI now provides a majority of spectroscopic calibration at redshifts below $z\approx1.2$. \rev{The dashed line at $z = 1.7$ provides a coarse upper bound for the detection limit of the [OII] feature on silicon based detectors, and thus the upper redshift limit anticipated by these samples.}}
    \label{fig:cellmedzs}
\end{figure}

The gains that have been made in constraining the \textit{p(z|c)} can be seen in \rev{the other panels of }Fig. \ref{fig:zmedians}, where we examine the additional color coverage of bright, low-redshift cells that were under-observed in the COSMOS field in the past. The red and magenta regions in Fig. \ref{subfig:countdiff} depict regions where DESI dominates the redshift information in the map by spectroscopic count. Cosmic variance and simple under-sampling can distort the broader color-redshift relation \rev{we measure by biasing populations of galaxies within a given SOM cell (see an extended discussion in App. \ref{app:som_verif})}. DC3R2 and DESI together provide much higher confidence in the distributions of redshifts at these magnitudes and colors. This is exemplified in Fig. \ref{fig:cellmedzs}, which depicts the number of cells in the map at each redshift with spectroscopic coverage. We see a dominant DESI + KV contribution in the BGS and ELG redshift regimes (the two peaks), which sharply drops off at $z = 1.55$ (becoming subdominant to COSMOS at $z > 1.375$), where the \rev{[OII]} line passes out of the optical range. DESI jointly with KiDS-VIKING calibrates \note{87\%} of the LSST / Euclid color-space (as a fraction of cells) at $z < 0.35$, \note{84\%} from $0.8 < z < 1.2$, and \note{77\%} overall for $z < 1.65$. While this is the overarching calibration extent of early DESI data in the KV fields, we can see that the higher coverage where DESI main survey targets supply galaxy spectra reveals that future efforts to span deeper magnitudes, and DC3R2-like efforts to bridge the spaces between main target categories could make real gains to fill more of the space approximately between $0.4 < z < 0.8$. For the full color space independent of redshift, the DESI-KV sample, inclusive of DC3R2, calibrates \note{56\%} of the space.\par
\begin{figure}
    \centering
    \includegraphics[width=\columnwidth]{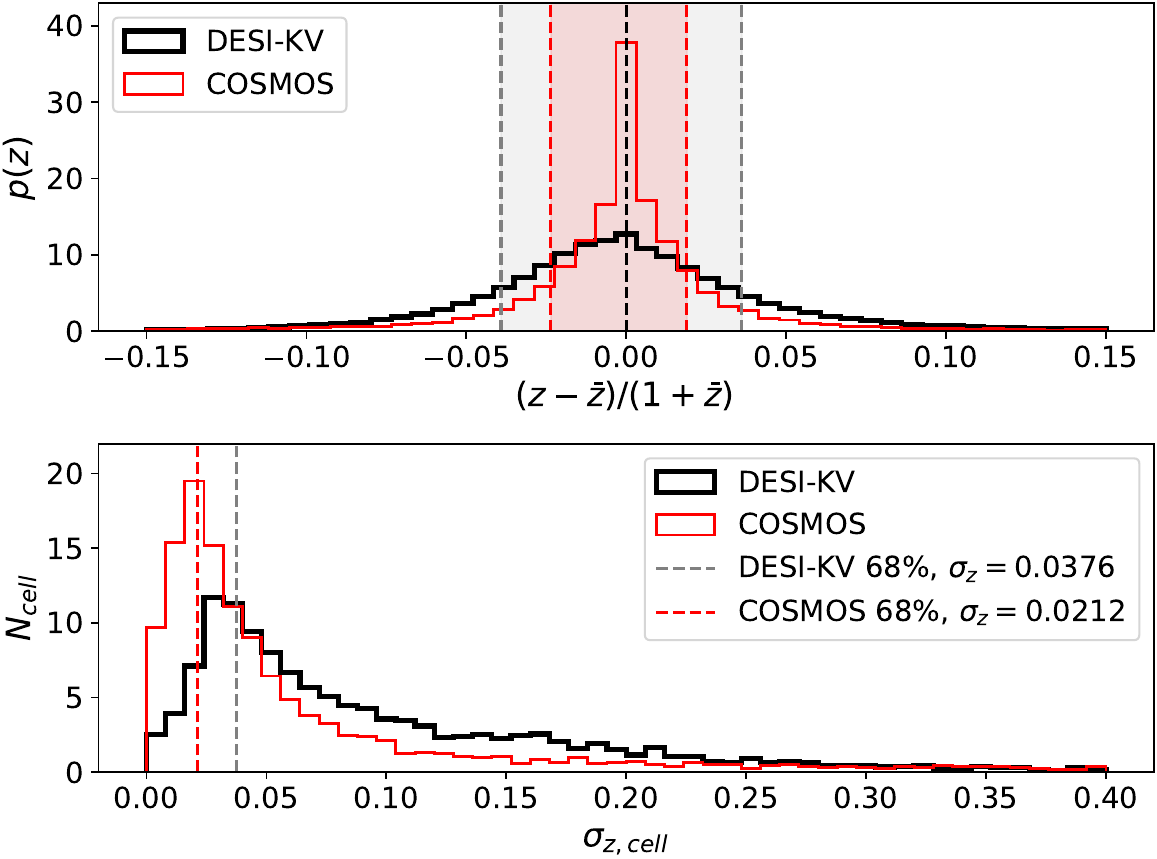}
    \caption{(Top) The distribution of redshifts across all cells as separation from mean redshift in a given cell and their respective $1\sigma$ region for DESI-KV and (for comparison) COSMOS. (Bottom) Distribution of the unbiased estimator for $\sigma_z$ cell widths across the SOM. The broader peak observed in this work is likely due to the shallower KiDS-VIKING wide field photometry that DESI spectroscopic redshifts are matched to. Noted in the legend are the shaded regions from the top panel.}
    \label{fig:cell_sigmazs}
\end{figure}

Similarly, DESI will produce better constraints on the $p(z|c)$ in the future, jointly with photometry of improved depth and with the full extent of the five year survey. With currently available KiDS-VIKING, we can see in Fig. \ref{fig:cell_sigmazs} that redshifts for a given color are more broad than in COSMOS- by a factor of \raisebox{0.5ex}{\texttildelow}1.7 - likely due to photometric scatter affecting cell assignment. The 68\%-region for the span of redshifts in a cell in aggregate, $(z_c - \bar{z}_c) / (1 + \bar{z}_c)$, with the DESI-KV sample is [-0.0392, 0.0361], yielding an approximate $\sigma_z = 0.0376$. In the bottom panel we can see how the distributions vary in individual cells, with the median of cells falling at approximately this aggregate (marked by the dashed lines). DESI provides a powerful data set, and if similar photometry to the COSMOS field were supplied for our spectra, we might anticipate our estimate of the variance in our overall redshift distributions statistically decreasing by a factor of 0.317 over our current reported, unweighted uncertainty (i.e. $\sigma_{z,\ \mathrm{COSMOS}}^2 / \sigma_{z,\ \mathrm{DESI-KV}}^2$). With an already substantial increase in number of spectra, deeper photometry would provide strong advantages. While the depth of the sample can be increased, the bounds on redshift range are fixed on the current instrument.


\section{Characterization of Spectroscopic Selection Biases on Redshift Calibration}
Ideal photometric redshift calibration for weak lensing surveys has to account not just for the dependence of redshift on observed color. Even at fixed color, magnitude and explicit and implicit selections of the spectroscopic or the weak lensing source galaxy sample can have an impact on the resulting redshift distribution (e.g. \citealt{gruen_2017,GruenNewman}). Testing or accounting for this requires a spectroscopic sample that simultaneously spans the color-space and the depth of the \rev{wider shape catalog for weak lensing}, and a full understanding of the selection process. Accounting for selection effects is difficult, but may be achieved via increasingly realistic simulations of the survey transfer function (e.g.~\cite{Myles_2021,Everett_2022} for weak lensing source galaxy selection accounted for in redshift calibration). For future surveys, spanning the depth will no longer be possible across all colors on account of the required spectroscopic exposure times. A sub-cell calibration that accounts for magnitude variation may become necessary at this stage, and the accuracy of this calibration will heavily depend on the depth and multiplicity of the available spectroscopic redshifts.

Here we investigate the impact of these systematics on redshift calibration with the DC3R2 spectroscopic sample. Section~\ref{sec:bias} tests the effects of magnitude cuts and spectroscopic selections on the redshift distributions of realistic Stage-III lensing survey redshift bins. Sections \ref{sec:magdepend} and \ref{sec:scatter} explicitly check for trends in redshift as a function of magnitude at fixed color, and the impact of noise in observed colors on those. Finally, Sections~\ref{sec:futuresurvey} and \ref{sec:interp} compare the systematic effects found to the requirements on future surveys.

\subsection{Impact on Calibration of Redshift Bins}
\label{sec:bias}
\begin{figure*}
    \centering
    \caption{A simple redshift calibration scheme for the G09, G12, G15 fields of KiDS-450 using DESI + DC3R2 spectroscopy reveals that applying selection effects on the spectroscopic catalog can significantly underestimate the mean redshift of a photometric sample of the same observed color.}

    \begin{subfigure}[t]{0.425\textwidth}
        \centering
	\includegraphics[width=1.08\textwidth]{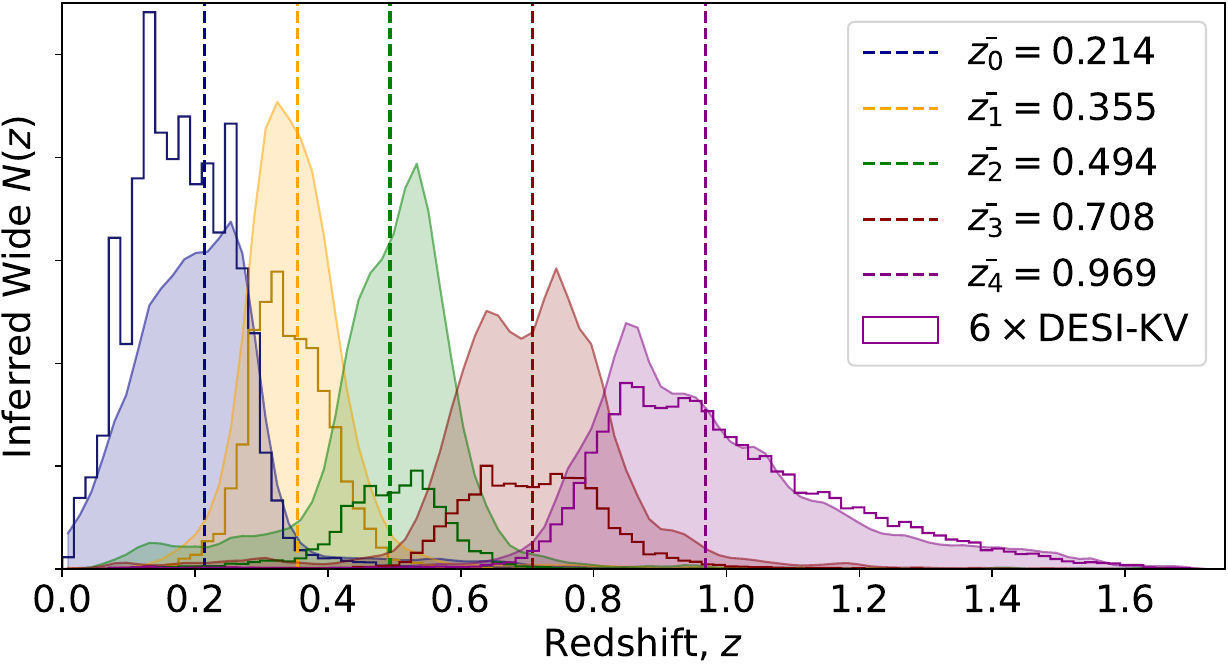}
        \caption{DESI spectroscopic redshifts (histogram outlines) and redshift distributions of photometric bins from re-weighting these by KiDS-450 color-space abundance (shaded). Shown are five tomographic bins of equal galaxy count selected by KV color after weighting (i.e. the bins from inferred redshift have equal size). The dotted lines represent the mean of the weighted bin, depicted in the legend. Estimated redshift distributions are smoothed by a Gaussian kernel for visualization purposes.}
        \label{fig:tomobins}
    \end{subfigure}
    \hspace{0.08\textwidth}
    \begin{subfigure}[t]{0.425\textwidth}
        \centering
	\includegraphics[width=1.08\textwidth]{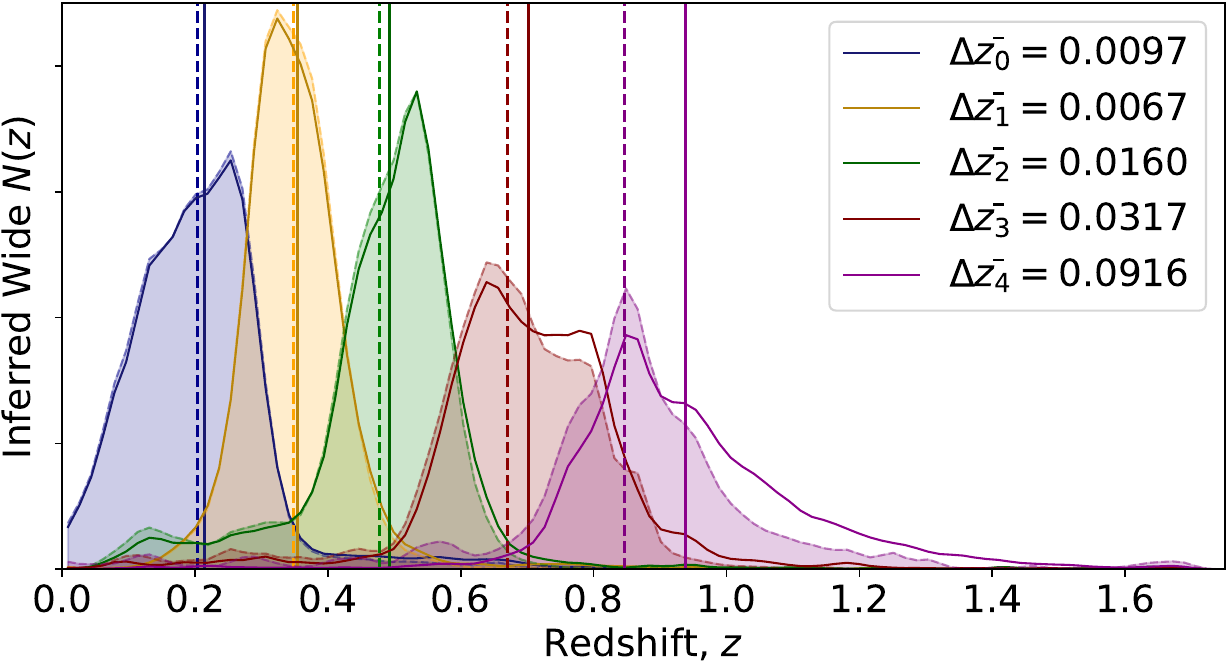}
        \caption{Inferred redshift distributions of KiDS-450 using a representative spectroscopic calibration sample (unfilled distributions) with the mean redshift of each bin indicated by the solid vertical line. When using a calibration sample with an additional selection of $\mathrm{MAG\_GAAP\_Z} < 21.0$, the inferred distributions (shaded) are altered and the mean redshift of each bin (dotted vertical line) is biased. The bias in mean redshift for each bin is noted in the legend, showing the substantial effect on the inferred redshifts of a photometric sample if the spectroscopic calibration sample is biased to brighter galaxies.}
        \label{fig:tomomag}
    \end{subfigure}
\end{figure*}

Selection effects in spectroscopic samples will distort crucial estimates for weak lensing surveys, like that of the mean redshift per tomographic bin. To explore these selection effects we use the KiDS-450 galaxy abundances in our SOM cells to infer redshift distributions for five KiDS-like tomographic bins. This procedure follows where,

\begin{equation}
    p(z| b\mathrm{, sel}) = \sum_{\mathrm{cell}\in b} p(z|c)\ p(c|\mathrm{sel})
\end{equation}
for a given selection, \textit{sel}, and bins, \rev{$b$}, comprised of cells, $c$. The first term amounts to the distribution of spectroscopic redshifts in a given cell and the second amounts to a weighting factor that relies on the abundances of galaxies in the calibrated sample (here, KiDS-450). The resulting $n(z)$s are depicted in Fig. \ref{fig:tomobins}. \rev{We create these bins by sorting cells covered by DESI spectroscopy by their median redshift. The lowest redshift cells are assigned to the first tomographic bin, until the number of KV galaxies with shapes occupying those cells reaches one fifth of the overall sample. We continue in this fashion, associating roughly equal numbers of calibrated galaxies with each bin, $b$. Note, that we can see in Fig. \ref{fig:tomobins} that this binning does not place equal numbers of \textit{calibrating} galaxies, i.e. spectroscopic redshifts, in each bin. }

We then perform a set of tests on the impact of selection biases by applying further selections to the spectroscopic sample and determining how that changes the estimated  $n(z)$s. Table \ref{tab:deltaz} summarizes the resulting change in mean redshift for the tomographic bins. 

\begin{table*}
    \centering
    \begin{tabular}{c|c|c|c|c|c}
    \hline
    Selection & $\Delta \bar{z_0}$ & $\Delta \bar{z_1}$& $\Delta \bar{z_2}$& $\Delta \bar{z_3}$ & $\Delta \bar{z_4}$\tabularnewline
    \hline
    $Z < 21.0$ & 0.0097 $\pm$0.0057& 0.0067 $\pm$0.0027& 0.0160 $\pm$0.0041& 0.0317 $\pm$0.0058& 0.0916 $\pm$0.0068\\
    $\Delta \chi^2 > 40$ & 0.0007 $\pm$0.0057& 0.0010 $\pm$0.0027& 0.0005 $\pm$0.0040& 0.0015 $\pm$0.0058& 0.0045 $\pm$0.0119\\
    $dz < 0.113 $ & 0.0182 $\pm$0.0064& 0.0086 $\pm$0.0030& -0.0115 $\pm$0.0040& -0.0036 $\pm$0.0057& 0.0034 $\pm$0.0124 \\
    $w(b,c,s)\ \mathrm{scheme}$ & -0.0041 $\pm$0.0055& -0.0062 $\pm$0.0031& -0.0152 $\pm$0.0042& -0.0155 $\pm$0.0055& -0.0033 $\pm$0.0130\\
    $Z < 21.0, w$ & 0.0016 $\pm$0.0054& 0.0014 $\pm$0.0031& 0.0031 $\pm$0.0042 & 0.0323 $\pm$0.0055 & 0.0642 $\pm$0.0071\\
    \hline
    \end{tabular}
    \vspace{0.05 in}
    \caption{Consolidation of the shifts in mean redshift inferred for a tomographic bin with KiDS-VIKING-like cell abundances if different selections are made on the spectroscopic sample. $\Delta \bar{z}_i$ is always the mean redshift of the new selection subtracted from the fiducial mean. The final row is a repeat of the first magnitude cut with the weights applied both before and after the selection. Error bars are produced by bootstrapping cells with replacement in the selection.}
    \label{tab:deltaz}
\end{table*}

Every comparison ensures that the color selection for each bin does not change as the spectroscopic selection effect is applied, i.e.~is made for bins defined by the same cell envelope both before and after the additional spectroscopic selection. This means that if the selection eliminates certain SOM cells from spectroscopic coverage, we use the estimated mean redshift of the same reduced set of cells even for the fiducial sample in order to isolate the effect of selection bias at fixed color.

The spectroscopic selection effects we examine are:
\begin{itemize}
    \item \textbf{Magnitude - ($\rm{MAG\_GAAP\_Z} < 21.0$)} For  $dz/dm=0$ and a well populated color space, a magnitude cut would introduce no change in mean redshift. In the presence of significant $dz/dm>0$, however, a magnitude cut would bias the inferred redshift towards a lower mean value. This operation and the result is demonstrated in Fig. \ref{fig:tomomag}. We indeed find a small ($|\Delta z|<0.01$) impact of the magnitude cut on the mean redshift inferred for the two lowest redshift bins, but an impact that exceeds current calibration requirements on the mean redshift inferred for the higher redshift bins, up to almost $\Delta \bar{z} = 0.1$. This cut changes the mean magnitude in each bin, and the spectroscopic redshift counts according to Table \ref{tab:magcut}. As this cut was chosen to be moderately bright for demonstration, we also explore how less severe selections affect this metric in Fig. \ref{fig:spec_bright_cut} as half magnitude steps down from the KV limiting magnitude in the i-band. This demonstrates the importance of representative spectroscopic redshifts in our toy model, which is constructed in a way that fundamentally differs from DES, KiDS, and HSC as our SOM is not trained on a magnitude in addition to its colors.

\begin{table}
    \renewcommand{\arraystretch}{1.5} % Default value: 1
    \centering
    \begin{tabular}{c|c|c|c|c|c}
    \hline
    Bin & 0 & 1 & 2 & 3 & 4\tabularnewline
    \hline
    $(\overline{Z_{\rm{mag}}})_{\rm{\ fid}}$ & 19.54 & 19.80 & 20.20 & 20.78 & 21.45 \\
    \rev{$(\Delta \overline{Z_{\rm{mag}}})_{\mathrm{\ cut}}$} & -0.163 & -0.197 & -0.308 & -0.499 & -0.949 \\
    $\Delta N_{\rm{spec}}/N_{\rm{spec,\ fid}}$ & 0.020 & 0.033 & 0.113 & 0.273 & 0.821 \\
    \hline
    \end{tabular}
    \vspace{0.05 in}
    \caption{Changes in properties of a given tomographic bin when \rev{their calibrating spectroscopic sample is subject to $\rm{MAG\_GAAP\_Z} < 21$}, describing the decrease in mean magnitude \rev{of the calibrating galaxies}, and the fraction of spectroscopic calibrating galaxies cut from each bin. \rev{Higher redshift bins are dominated by fainter spectra and suffer the largest population shifts.}}
    \label{tab:magcut}
\end{table}
\begin{figure}
    \centering
    \includegraphics[width=\columnwidth]{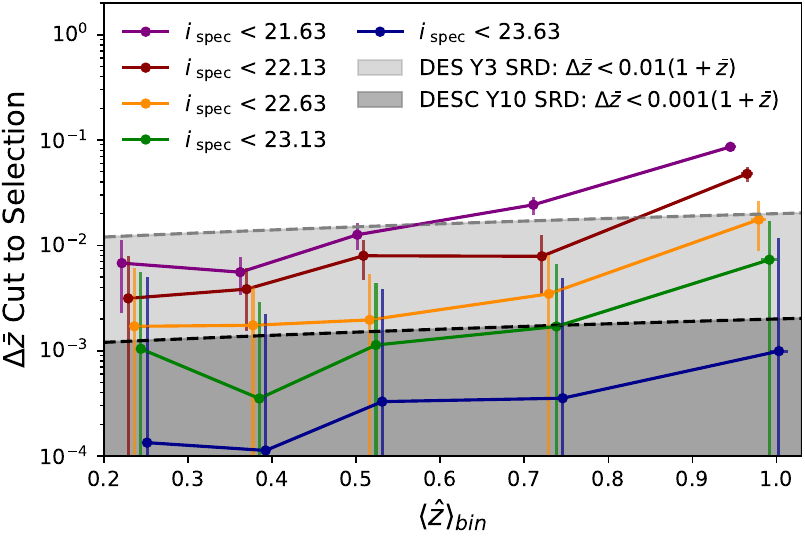}
    \caption{Lack of representative spectra for an inference biases the mean redshift of a bin for our sample matched to KV. Depicted are the shifts in mean redshift for our tomographic bins should the calibrating spectroscopic galaxies be preferentially brighter than the source galaxies. We show several choices of magnitude cut incrementally brighter than the KiDS limiting magnitude of $i = 23.63$, and compare that to the Rubin \rev{Science Requirements Document (SRD)} \citep{lsstreqs2021}. Small offsets have been applied to the mean redshift per bin, $\left<\hat{z}\right>_{\mathrm{bin}}$, for visualization.}
    \label{fig:spec_bright_cut}
\end{figure}
    \item \textbf{Quality Flags - ($\Delta \chi^2 > 40$)} As a consequence of enforcing a more severe confidence in the fitted model, this will remove spectra in an SED-dependent way from the ELG and LRG samples where, by default, some objects with smaller $\Delta \chi^2$ are considered confident redshifts. When implemented, less then 2\% of spectroscopic galaxies are removed.The impact of this stricter selection is small ($|\Delta z|<0.005$) but coherent across all redshift bins, i.e.~biases all redshift distributions towards a lower mean by preferentially removing higher redshift galaxies from the sample within a given cell. In the highest redshift bin, where the fraction of ELG targets contributing to the calibration is large, the effect exceeds the redshift calibration requirements of future weak lensing experiments (cf. also \citealt{Hartley_2020} for the impact of selection based on conventional redshift confidence flags).
    \item \textbf{Removal of Cell Outliers - } Galaxies with large deviation from the median redshift of an ensemble of similar color are more likely to be outliers of various types, e.g. due to blending, AGN light, or redshift determination errors. We define a sample of outliers based on the criterion
        \begin{equation}
            \label{eq:outlier}
            dz = \frac{|z - \rm{median}(z)_{cell}|}{(1+\rm{median}(z)_{cell})}  \rev{>} 3\sigma_{z,{\rm all}} \; ,
    \end{equation}
    where $\sigma_{z,all}$ is the aggregate standard deviation in cells found in Sec. \ref{sec:colorspace}, from which $3\sigma_{z,all} = 0.113$. The selection corresponds to the central 99.7\% of redshifts if these were to follow Gaussian distributions of color-independent width at any fixed color and thus would reject only the most egregious outliers. In practice, these distributions are not Guassian and this selection cuts 5.1\% of objects. The selection is relative to the median rather than the mean to robustly deal with cells that are undersampled or heavily affected in their mean redshift by outliers. The impact of removing the outlier population defined this way is maximally of order $\Delta z \approx 0.02$, and has reduced impact in the highest redshift bins. While a number of these outliers can be attributed to photometric scatter across degenerate regions in the SOM (neighboring cells with large separation in redshift), others may be real examples of broad or bimodal cell distributions and ought to be examined with visual inspection.

    \item \textbf{Applying Weights, $w$ - } Spectroscopic redshifts are weighted according to the scheme described in Section \ref{sec:weights}, to account for prioritization of spectroscopy for targets of certain morphologies, colors, and SED-types that are not necessarily representative. As the weights can be somewhat noisy in under-sampled regions of color space (in targeting), additionally this comparison is restricted to cells where the resulting shift in  $\bar{z}_c$ from applying the weights is small. While this will result in an underestimation of the true weighting effect for all galaxies, these are the regions where weights can be applied confidently due to the spectroscopic and targeting counts. We apply a cell envelope selection where $|\bar{z}_c -\bar{z}_{c,w}|<0.08 $, which retains > 90\% of cells \rev{and eliminates the sparsely occupied, and noisiest cells}. Note that the effect of this scheme in Table \ref{tab:deltaz} is always to lower the mean redshift in a given bin. The impact is maximally $\Delta z \approx 0.01$ in the most sparsely populated spectroscopic bins. Additionally, if these weights are applied in the same way to the magnitude selection ($Z < 21$), we see in Table \ref{tab:deltaz} that they mitigate the bright spectroscopic bias in high bins, but do not eliminate it.
\end{itemize}

\subsection{Magnitude Dependence of Redshift at Fixed Color}
\label{sec:magdepend}
Past analyses on photometry and spectroscopy in the COSMOS field have shown the magnitude dependence of redshift at fixed color cell to be small and well described by a linear behavior with $ dz/dm \approx 3\times 10^{-3}$ \citep{Masters_2017}. The equivalent measurement made in the KV data with DESI redshifts is depicted in Fig. \ref{fig:dzdm}. Here we examine the relation of differences in \revtwo{VIKING} Z-band magnitude and redshift between pairs of galaxies that occupy the same color-cell in the SOM. A cell with $n$ galaxies contributes  $n(n-1)/2$ data points to Fig. \ref{fig:dzdm}. We see a linear relationship, with a large amount of scatter. The raw slope measured, \note{$dz/dm = 0.0250 \pm 0.0009$}, is an order of magnitude larger than that reported in previous studies (see App. \ref{app:dzdm_slope} for methodology). This ought not to be taken at face value, and Section \ref{sec:scatter}  (Fig. \ref{fig:photscatter}) explores photometric scatter as the largest source of bias in this measurement, as well as the chief difference between the data set used in this study and that of previous ones. Furthermore, \ comparisons of $dz/dm$ will be affected to lesser degrees by the other second order effects discussed in Section \ref{sec:verification}. 

\subsection{Systematic Uncertainty due to Photometric Scatter}
\label{sec:scatter}

With the depth of our SOM resembling that of future deep surveys, it could be problematic that we use a relatively shallow photometry in KiDS-VIKING to assign our spectroscopic galaxies. Photometric scatter enters our calculation through perturbations in the cell assignment, which comprises our definition of fixed color. \rev{Galaxies with larger flux uncertainties tend to be fainter, which is also correlated with higher redshifts. While the direction of scatter will be impacted by the color topography of our individual SOM, we can broadly expect faint, high redshift galaxies to scatter more often into brighter, lower redshift cells and induce a strongly positive $dz/dm$. We can imagine photometric scatter as an asymmetric smearing of the redshift distributions in a given cell.}
\par
\begin{figure*}
    \centering
    \caption{Plots depicting the systematic from photometric scatter induced in $dz/dm$ (left) and the raw measurement from the joint DC3R2-KV sample (right), where each data point is the difference between a unique pair of galaxies occupying the same SOM color-cell in magnitude and redshift. The best-fit slope (dashed red) is fit from a collection of medians and $\sigma$ (pink) in magnitude slices. If reported photometric errors are accurate, this systematic dominates our measurement.}
    \label{fig:dzdmbig}
    \begin{subfigure}[t]{0.425\textwidth}
        \includegraphics[width=1.12\textwidth]{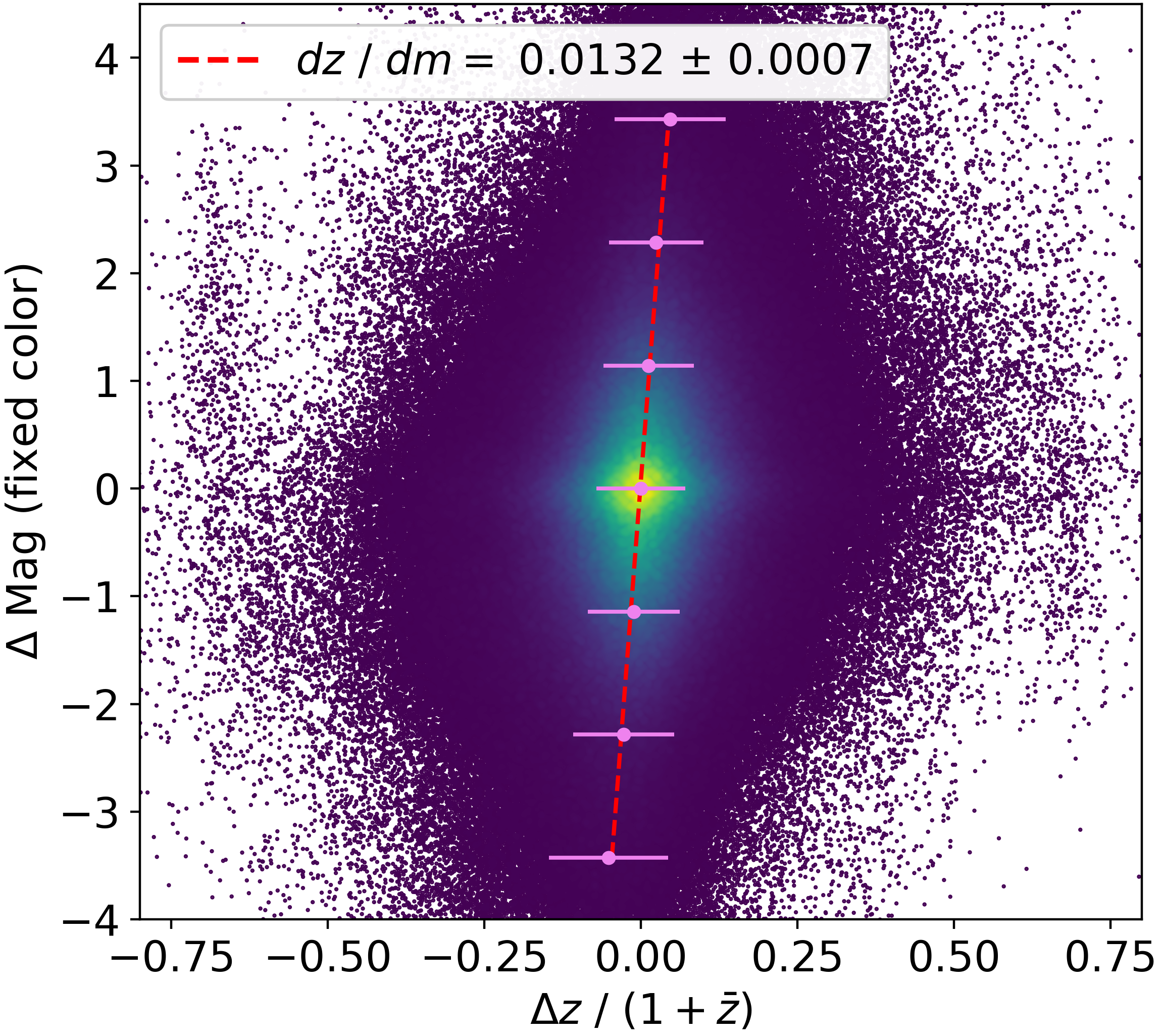}
        \caption{Systematic induced in $dz/dm$ by KiDS-VIKING photometric errors in conjunction with our given SOM resolution for DC3R2 spectroscopic redshifts at fixed color, with change in magnitude in \revtwo{KiDS-VIKING} MAG\_GAAP\_Z. \rev{Sec. \ref{sec:scatter} describes the procedure of scattering galaxies by their reported errors to perform this measurement.}}
        \label{fig:photscatter}
        \end{subfigure}
     \hspace{0.08\textwidth}
    \begin{subfigure}[t]{0.425\textwidth}
	   \includegraphics[width=1.12\textwidth]{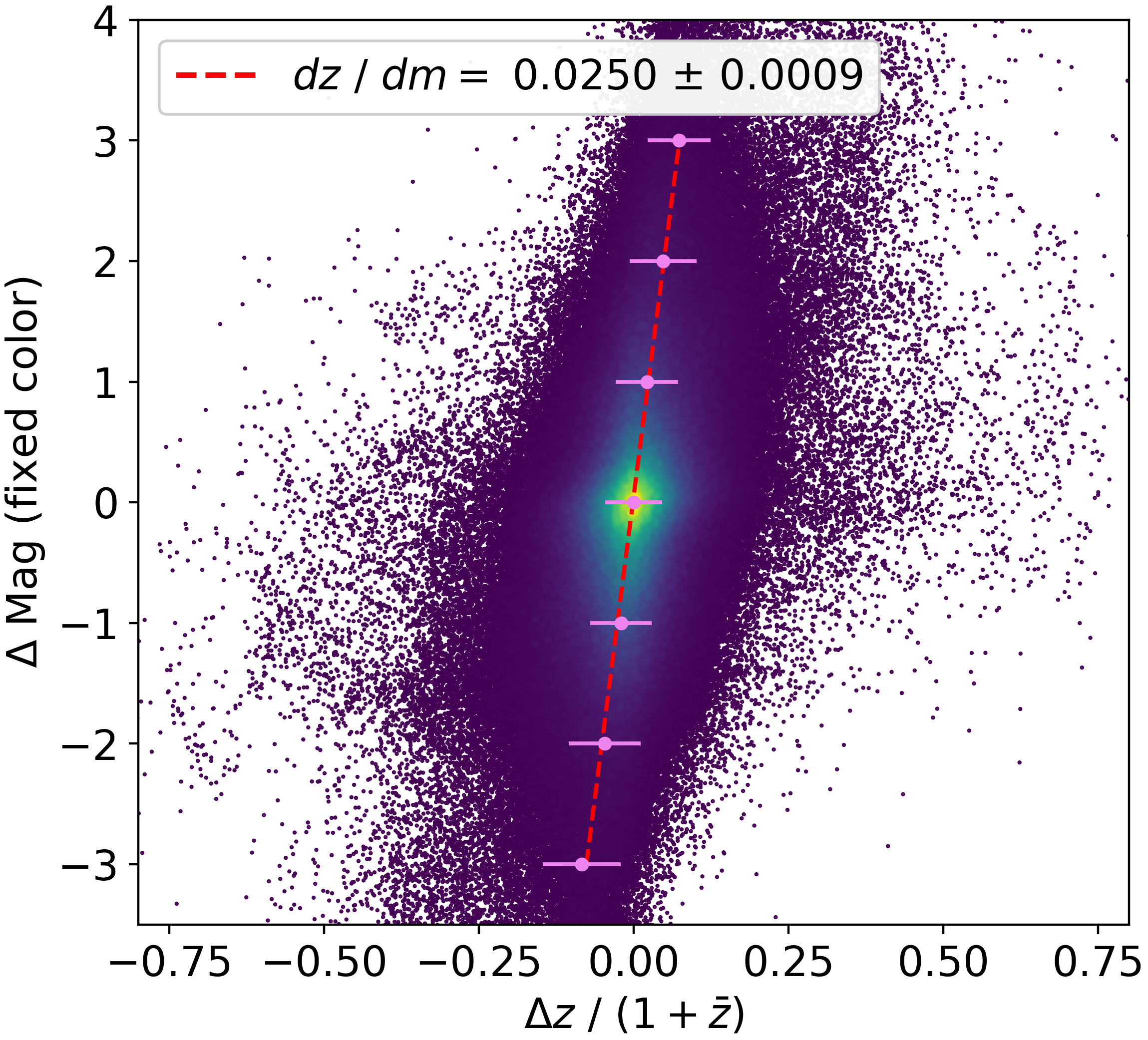}
        \caption{\rev{Total measured change in redshift with magnitude (Z-band) at fixed color} as observed in the \rev{shallow} KiDS-VIKING photometry is dramatically affected by photometric scatter \rev{in contrast to previous studies on deeper COSMOS field photometry}.}
        \label{fig:dzdm}
    \end{subfigure}
\end{figure*}
\begin{figure*}
    \centering
    \caption{Color dependence of $dz/dm$ in the raw data (left) and corrected for the contribution of photometric scatter to the overall slope (right).}
    \begin{subfigure}[t]{0.30\textwidth}
        \includegraphics[width=1.12\textwidth]{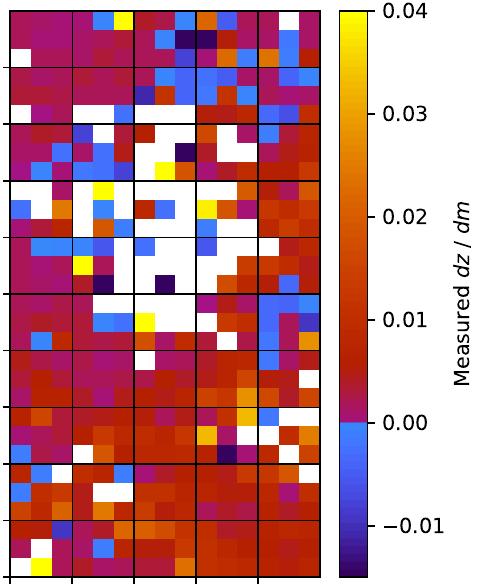}
        \caption{Raw measurement of change in redshift with magnitude at fixed color, $dz/dm$ across the SOM, deresolved for statistical power into 5 x 5 super cells.}
        \label{fig:dzdm_raw_som}
        \end{subfigure}
     \hspace{0.08\textwidth}
    \begin{subfigure}[t]{0.30\textwidth}
	   \includegraphics[width=1.12\textwidth]{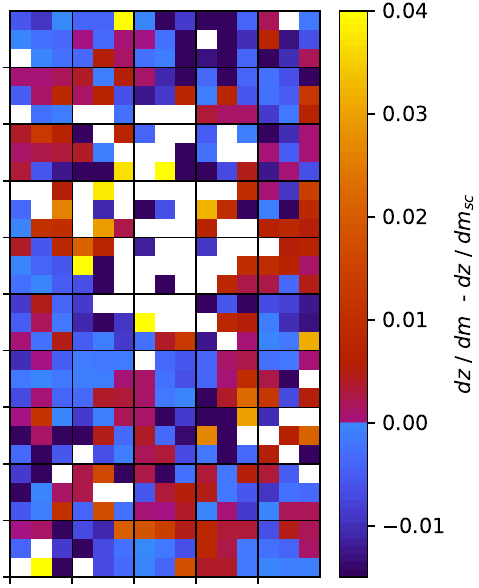}
        \caption{Intrinsic change in redshift with magnitude at fixed color, $dz/dm$ as observed across the SOM, corrected to first order by subtracting off the contribution due to photometric scatter (i.e. $dz/dm_{\rm raw} - dz/dm_{\rm sc}$).}
        \label{fig:dzdm_corrected_som}
    \end{subfigure}
    \label{fig:dzdm_cell}
\end{figure*}

A potential systematic effect of this photometric scatter on the measured slope can be explored with a test described here. We can perform a direct measurement on the slope introduced by photometric noise by (1) removing any existing $dz/dm$ from our data by randomly shuffling all spectroscopic redshifts in a given SOM cell, thus nulling $dz/dm$ while preserving the $p(z|c)$, (2) applying a random Gaussian draw of the width of the flux error reported in the catalog for each band, thus perturbing measured colors. (3) Reassigning galaxies to the SOM based on these perturbed fluxes, we repeat our measurement of $dz/dm$. Additionally, we can (4) iteratively reshuffle and perturb many times to reduce statistical uncertainty in the measurement from the limited sample size. The only contributor to this final, measured $dz/dm$ will be photometric scatter, making it a measurement on this systematic that can be compared across photometric surveys of different depth. Using (4) to quintuple the number of points available to us for this measurement, we observe in Fig. \ref{fig:photscatter}, that \note{$(dz/dm)_{sc} = 0.0132 \pm 0.0007$} for the reported KV errors associated with our DESI targets in this SOM. The slope \rev{induced} this way can be measured across color space, i.e. across the SOM. This color-dependent effect is subtracted off in the second panel of Fig. \ref{fig:dzdm_cell}, leaving an estimate of an intrinsic $dz/dm$ that is corrected at first order, at least under the assumption that the reported flux measurement errors are accurate. While individual cells carry significant large intrinsic $(dz/dm)_c$ (order of $\approx 0.01$), the average across all cells is much lower at ${(dz/dm)_{c,avg.}} = -7\times10^{-4}$.

\subsubsection{Impact of photometric noise levels}
\label{sec:depth}
We can see in Fig. \ref{fig:photscatter} that the systematic measured for KiDS-VIKING-like error in a SOM of Euclid-like resolution is significant, \note{$(dz/dm)_{\rm sc} \approx 0.013$}, and the likely dominant contributor to our overall measured slope. With this effect in mind, we revisit the original measurement performed in \cite{Masters_2019} using the same field and similar photometric filters. Repeating the test of Section~\ref{sec:scatter} used to measure the impact of photometric scatter, but with the lower noise levels of the COSMOS photometry, we find that the original measurement had a contributing noise bias of $(dz/dm)_{\rm sc} \approx 0.003\pm0.0004$. This is similar to the measured slope in \cite{Masters_2019}, and means that the intrinsic $dz/dm$ in that sample is roughly consistent with zero. With the limitations of the coarse photometric scatter test we have applied we do not claim to have a measurement of the \textit{intrinsic} $dz/dm$ in COSMOS to better than $0.003$ as a result.

Also worthy of note is that doubling the photometric error for KV associated DESI redshifts produces a \note{$(dz/dm)_{\rm sc,\ 2} = 0.0256 \pm 0.0010$}, which is larger than the observed raw slope in Sec. \ref{sec:scatter}. Thus, in case the photometric error in the KV catalogs should be underestimated at levels that have been reported in other studies doing source injection into survey images \citep[e.g.]{Everett_2022}, it could be that the average intrinsic slope of the DC3R2 sample is indeed consistent (across the full sample, not simply cell-by-cell) with zero as well.

As noise in the KiDS-VIKING photometry introduces a large systematic effect on our desired measurement, it is worth exploring what photometry is needed to drive down the error and improve constraints on $dz/dm$. Given the uncertainty on our estimation of the systematic effect of photometric noise, the hypothesis that $dz/dm\approx 0$ seems to be consistent with current data. Yet improved photometry such as that from HSC \citep{Aihara2018} could allow to use DESI spectra to better effect. \rev{With a Gaussian-draw, background limited error model based upon the effective exposure times and limiting magnitudes of \citealt{Aihara2018} (see their Table 8) we generate} HSC-like flux errors for \textit{grizy} to augment KiDS-VIKING \textit{uJHKs}. We find that $(dz/dm)_{\rm sc} \approx 0.008$, which halves the effect, though does not eliminate it. In the limit of noiseless optical (\textit{ugriz}), we still find $(dz/dm)_{\rm sc} \approx 0.004$. We can imagine the LSST scatter with these objects would be comparable to this value, without improved NIR or IR follow up. Note that this is still larger than the slope found with the COSMOS photometry that includes the very deep UltraVISTA NIR data, demonstrating both that such small systematic errors are achievable in principle, and that they rely on deep NIR that is difficult to achieve with currently operating instruments.

\subsection{Future survey requirements}
\label{sec:futuresurvey}

The tests presented above have shown limitations to how well we can constrain the trend of mean redshift with magnitude at fixed color in the presence of noisy photometry. Here, we connect this to requirements for how well do we need to know $dz/dm$ for future, stage IV weak lensing surveys. 
%If we have established that the best we can do with current NIR photometry and DESI Y1 data is $\Delta(dz/dm) \approx 0.004$, then we ought to prescribe a rationale for why this matters. 
We can approximate the redshift calibration error due to imperfectly known magnitude dependence as 
\begin{equation}
    \frac{\Delta z}{(1+\bar{z})} \approx \Delta(dz/dm)\times (\bar{m}_{\rm wide} - \bar{m}_{\rm spec}),
    \label{eq:zerrlsst}
\end{equation}
where $\Delta(dz/dm)$ is our error in the magnitude dependence of redshift at fixed color, $\bar{z}$ is the mean redshift of the sample being calibrated, and $(\bar{m}_{\rm wide} - \bar{m}_{\rm spec})$ is the offset between the mean magnitudes of wide field observed galaxies and the spectroscopic sample used to calibrate them. 

The requirement for final Rubin analyses is expected to be $|{\Delta z}|/{(1+\bar{z})} \approx 0.001$ \citep{lsstreqs2021}. With the limiting $i$ band magnitude of final Rubin data and the mean magnitude of the current DESI sample for $\bar{m}_{\rm wide}=26.3$ and $\bar{m}_{\rm spec}=21.1$, respectively, we find that we need to determine $|\Delta(dz/dm)| \le 0.0002$. If we assume the $\Delta(dz/dm) \approx 0.004$ that was determined in Section~\ref{sec:depth} to be the limit obtainable with current NIR photometry and deep Rubin observations, our needs and capabilities are at odds.

If instead of considering the faintest galaxies used by Rubin we estimate the sample's mean magnitude, using the slope of the luminosity function from \cite{gruen_2017}, observed to the limiting i-band magnitude for LSST, we obtain $\bar{m}_{\rm wide}=\bar{i}_{\rm Rubin}\approx 24.90$. This provides a more realistic, less conservative requirement for  $\Delta(dz/dm) \le 0.0006$, which is still stricter than all our estimates for the effect of photometric scatter on this slope by almost an order of magnitude. %If we are to meet photometric redshift requirements for future surveys with an expectation of low redshift dependence on magnitude, we will have to model photometric scatter as a systematic effect.

\subsection{Interpretation}
\label{sec:interp}

The previous sections studied how selections on the spectroscopic calibration sample provided here may impact the redshift distributions estimated using that sample. We performed these tests using a simple redshift calibration scheme that assumes that at a fixed observed color, the distribution of true redshifts of a photometric and a spectroscopically selected galaxy sample are identical. We note that none of the recent, Stage-III, analyses have relied on a redshift calibration scheme that was quite this simple, and hence the biases we identify are not expected to be present at the level found here in recent analyses. For example, Fig. \ref{fig:spec_bright_cut} depicts a case where the calibrating spectra are up to two magnitudes brighter than the limiting magnitude of the sample, whereas the formal KiDS analysis made use of a variety of spectroscopic sources that spanned the depth of their photometric sample \citep{van_den_Busch_2022}. Similarly DES did not strictly apply a spectroscopic selection and made use of narrowband photometric redshifts where there were no representative spectroscopic redshifts \citep{Myles_2021}. Despite this, extrapolation of redshift calibration samples to photometric objects of similar color but fainter magnitude is likely required for future, deeper photometric surveys, and hence it is useful to study the potential pathways for bias in such an application of our sample. Both the DES and KiDS SOMs (\citealt{Myles_2021,Wright_2019}) are trained on colors \textit{and} a magnitude (or luptitude) -- and DESC will be unable to do this without throwing away a substantial part of their sample. Hence the SOM for this analysis is matching spectroscopic \textit{z}s to galaxies based on their colors alone.

A necessary condition for a non-zero bias in such a redshift calibration scheme is that the expectation value of redshift depends on properties relevant for the spectroscopic target selection and measurement success other than just a galaxy's observed color. The most salient such effect we identify is via a trend of mean redshift with observed magnitude at given observed color, $dz/dm$. We find this slope to be much steeper than reported in previous studies. The cause for this steepness is an effect of the larger photometric noise in the colors measured for our target sample, rather than a large dependence of redshift on magnitude at fixed \emph{true} color. Most of the imaginable selection effects are related to a galaxy's magnitude, and hence the non-zero $dz/dm$ propagates into several of the bias tests we perform and merits further interpretation.

The bias imposed by photometric scatter on $dz/dm$ varies with color, as seen in Fig. \ref{fig:dzdm_cell}, and potentially also with other selection choices. The scatter itself is asymmetric as it does not shift objects into neighboring cells isotropically in the map, and will induce a larger $dz/dm$ in areas with large color-redshift degeneracies (i.e. a small error in color could lead to a large offset in the mean local redshift). Our ability to understand the intrinsic $dz/dm$ present in the data depends heavily on the quality of our photometry and/or our ability to correctly model photometric measurement errors. As seen in the right panel of the same figure, the intrinsic $dz/dm$ is similarly color dependent, but just as likely to be negative as positive (indicating that it is likely a noisy measurement). 

Our method to constrain this bias relies on the reported photometric errors.  However, if these errors were to be misestimated by order unity, potentially in a color dependent way, the entire measured slope could be the result of this systematic, as discussed in Sec. \ref{sec:depth}.  Literature analyses have demonstrated that the deviation between observed photometry and the true fluxes of a galaxy requires realistic image simulations processed by photometric pipelines to estimate, and reported errors are frequently an underestimate at levels that indeed reach factors of two (e.g. \citealt{Huang_2017,Everett_2022}). No literature study exists on the accuracy of reported KV GAAP flux errors specifically, but the relative linearity of the algorithm and preliminary studies on image simulations \citep{li2022} imply that the misestimation of the error in GAAP is non-zero but not a factor of two.
\\
\\
Consequently, Sec. \ref{sec:bias} demonstrates that selection effects in magnitude suffer from $dz/dm$ dependence significantly in higher redshift bins, where objects tend to be fainter, and this emphasizes the role that deep photometry, in contrast to merely multi-band coverage, plays when making maximal use of spectroscopic redshifts. We can see with the $\sigma_z$ cut, on the narrowness of the redshift distribution for a given cell, that potential outliers in a given cell that arise either from photometric scatter or a misattributed redshift can have dramatic effects on the redshift distribution. Since this effect is largest in the lowest redshift bin--where objects are brightest--we might suspect that these outliers are dominated by cell-to-cell photometric scatter.  Cutting these cells does further limit the color space of a given weak lensing analysis, and ought to be avoided where possible. For this reason, visual inspection works like \citet{VI_paper} and careful selection of the redshift sample are crucial to future analyses.
\\
\\
It has been established (a) that LSST will require a very accurate measurement of $dz/dm$ if calibrated by DESI alone (Sec. \ref{sec:futuresurvey}), and (b) that even with perfect optical photometry the photometric scatter contribution to $dz/dm$ with existing DESI spectra is larger than the acceptable error by almost an order of magnitude (Sec. \ref{sec:depth}). The strategies to account for this in future weak lensing redshift calibration efforts can be threefold:
\begin{itemize}
    \item \textbf{Deeper Photometry} : Despite (b), specifically improving NIR/IR measurements will decrease the systematic contributor to $dz/dm$ and potentially allow for a measurement that meets photo-\textit{z} requirements.
    \item \textbf{Deeper Spectroscopy} : Deeper spectra will reduce a given survey calibration's dependence on $dz/dm$, mitigating how well known this slope has to be in (a). After DESI's current survey, it will be uniquely situated to push for deeper spectra.
    \item \textbf{Modeling} : Perhaps the most viable approach, one can take into account the effect of photometric scatter with appropriate modeling of the data. Future analyses could constrain the bias via forward modeling from a color space of true photometry. In this approach, the intrinsic and observed $dz/dm$ are folded into the inference alongside the systematic in a methodologically appropriate way, and the size of the systematic effect is no longer a limiting factor.
\end{itemize}

\section{Conclusions}

%This paper presents a catalog of DESI spectra and associated weights that attempt to make the sample representative at a given KiDS-VIKING color and a limiting magnitude. These will aid the redshift calibration of a large fraction of the lensing source galaxy samples of future photometric surveys. The main conclusions from this work include:
%Suggested shuffling: 
Photometric redshift calibration -- estimating the redshifts for galaxies where we only observe them through a collection of filters -- requires a thorough understanding of the color-redshift relation, which is very non-linear across the full galaxy population. This paper presents a catalog of spectra, the Dark Energy Spectroscopic Instrument (DESI) Complete Calibration of the Color-Redshift Relation (DC3R2) secondary target survey, designed to aid in the redshift calibration for a large fraction of the weak lensing source galaxy samples of future photometric surveys. The data includes associated weights that %allow to 
turn the survey samples (ELGs, LRGs, BGS) into one that is representative at a given $ugrizYJHK_s$ and limiting magnitude of 23.68 ($5\sigma$, $i$-band), using KiDS-VIKING as test-bed. We chose to select targets on this 9-band photometry in order to break redshift-type degeneracies, and KiDS-VIKING provided the most constraining data over this area. With this unprecedented quantity of DESI spectroscopic redshifts, we examine how it benefits future surveys and allows us to examine spectroscopic selection effects.

\begin{itemize}
    \item The DESI sample reported here calibrates the redshift distribution of roughly \note{56\%} of the galaxies in COSMOS via 230k spectroscopic redshifts. For the photometric colour space that will be visible to DESC and Euclid (approx. 98\% complete at $i = 25.3$), this sample corresponds to coverage of 6248 cells out of 11250. Approximately \note{41\%} of the full COSMOS color-space and galaxy population is calibrated from spectroscopic galaxies that are classed as DC3R2 targets (inclusive of overlap with main survey targets), and \note{4\%} from uniquely DC3R2 objects. 
    \item However, even though this sample provides an incredible quantity of high-quality spectra, we find that the combination of uncertain photometry and a variety of spectroscopic selection effects can produce substantial biases on the redshift inference for a given lensing redshift bin. We demonstrate that introducing a preference for brighter-magnitude calibration spec-\textit{z}s in the presence of photometric scatter effects or intrinsically high $dz/dm$ biases the mean redshift on the bin of order $\Delta z \approx 0.01$, especially for higher redshift bins (see Fig. \ref{fig:spec_bright_cut} for a breakdown). This effect is present even for the shallower test-bed data used here and will be exacerbated for fainter samples, as it enters the inference as an induced magnitude dependence of redshift at fixed color. Fewer photometric bands further worsens the effect, as breaking degeneracies becomes more difficult (e.g. the uncertainty in mean redshift for the year 3 HSC, \citealt{Rau23}, \textit{grizy} analysis was larger than that of KiDS-450, \citealt{Wright_2020} $ugriZYJHK_s$ by a factor of 1.25, even with KiDS having an additional bin).
    \item Results for this work are expressed in a color-space that is similar to past work on the \citet{Masters_2017} SOM, as our map is a transformation of this space into KiDS-VIKING colors. As demonstrated in Fig.~\ref{fig:spec_v_phot}, we recover very similar galaxy SEDs and redshifts per cell, but comparisons like that of Fig.~\ref{subfig:countdiff} ought to be taken with the knowledge that the colors and photometric noise levels between different surveys are not identical.
  \item Our analysis reveals a general agreement in the color-redshift relation with previous spectroscopic surveys that have explored this color-space. When accounting for effects induced by photometric noise, we also find agreement in the magnitude dependence of redshift at fixed color with the result of \citet{Masters_2017}.
    \item Photometry quality has an important role in redshift calibration. This study has found that photometric errors need to be well understood for modeling the color-redshift relation and especially magnitude dependent effects ($dz/dm$). In order to constrain this slope for future surveys we require either better photometry than expected, deeper spectroscopy, or improved analysis methodology (discussed in Sec. \ref{sec:interp}).  We find that the use of KV photometry in this map does not supply as strong a constraint as past data sets do on the slope $dz/dm$. However, close examination of the systematic $dz/dm$ induced by photometric scatter in a high resolution SOM has strengthened the case for the null hypothesis in this work and in past survey data sets. While $dz/dm \approx 0$ is potentially consistent with our data, deeper photometry or survey simulation will be needed to constrain the slope sufficiently for future weak lensing efforts.
    \begin{comment}
    \item As discussion in Sec. \ref{sec:bias}, weighting our spectra appropriately (in magnitude space as well as color-space) had an effect of \note{x}, which demonstrates that main survey targets despite their additional selections can be incorporated appropriately into future color-redshift relation inferences.
    \end{comment}
    \end{itemize}
The spectroscopic redshifts measured by the complete 5-year DESI survey will provide unparalleled support for future redshift calibration in weak lensing surveys. To fully leverage the powerful quantity of data discussed in this paper, more accurate photometry and deeper spectroscopic redshifts will be necessary to constrain the magnitude dependence of redshift at fixed color. Exploration into the capability of massively multiplexed spectroscopic instruments, like DESI, to attain redshifts of fainter sources than currently targeted is important. Looking ahead, future surveys will provide additional wavelength and sky area coverage that can improve photometric redshift calibration. Among these is a follow-on program to DC3R2 using the 4-metre Multi-Object Spectroscopic Telescope (4MOST; \citealt{4most}) that will observe targets across the same SOM used in this work, though more uniformly to redshift $\sim$1.55 \citep{4c3r2}. Together with deeper spectroscopic campaigns (e.g. DESI-II, \citealt{snowmass_desiII}) and campaigns including high-quality infrared spectroscopy, these data will form the basis for constraining a model of the galaxy population seen by deep photometric surveys, including its redshift distribution.
%%%%%%%%%%%%%%%%%%%%%%%%%%%%%%%%%%%%%%%%%%%%%%%%%%
\section*{Data Availability}

The data and code used to generate the figures in this paper will be made available with its publication\footnote{\url{https://zenodo.org/record/8328495}}. The redshifts for objects taken during SV are available in the DESI Early Data Release\footnote{\url{https://data.desi.lbl.gov/doc/releases/edr}} (see \citealt{desi_edr}). Visualization tools exist to explore the data across the color-space in browser and are available via github\footnote{\url{https://jmccull.github.io/DC3R2_Overview}}. \begin{comment}Additionally, a \note{value-added} catalog will be produced with cell assignments and weights for relevant objects. \end{comment}

\section*{Acknowledgements}
The authors thank several people for enabling this work, among them Mara Salvato, for providing the COSMOS collaboration spectroscopic catalog, and Hendrik Hildebrandt for providing an updated KiDS catalog covering the COSMOS region which allowed us to leverage spectroscopic measurements as well as perform crucial tests on our methodology. Other thanks go towards the wider DESI collaboration and the C3 working group for providing feedback and guidance at several stages of the process. 

%Funding Acknowledgements of main authors
JM received a Deutscher Akademischer Austauschdienst (DAAD, German Academic Exchange Service) fellowship and funding from the Bavaria California Technology Center (BaCaTeC) in support of this work. This research was also supported by the Excellence Cluster ORIGINS which is funded by the Deutsche Forschungsgemeinschaft (DFG, German Research Foundation) under Germany’s Excellence Strategy – EXC-2094-390783311.

%DESI Standard Acknowledgements
This research is supported by the Director, Office of Science, Office of High Energy Physics of the U.S. Department of Energy under Contract No. DE–AC02–05CH11231, and by the National Energy Research Scientific Computing Center, a DOE Office of Science User Facility under the same contract; additional support for DESI is provided by the U.S. National Science Foundation, Division of Astronomical Sciences under Contract No. AST-0950945 to the NSF’s National Optical-Infrared Astronomy Research Laboratory; the Science and Technologies Facilities Council of the United Kingdom; the Gordon and Betty Moore Foundation; the Heising-Simons Foundation; the French Alternative Energies and Atomic Energy Commission (CEA); the National Council of Science and Technology of Mexico; the Ministry of Economy of Spain, and by the DESI Member Institutions.

The authors are honored to be permitted to conduct scientific research on Iolkam Du’ag (Kitt Peak), a mountain with particular significance to the Tohono O’odham Nation.

%DESI Legacy Survey Acknowledgements
The DESI Legacy Imaging Surveys consist of three individual and complementary projects: the Dark Energy Camera Legacy Survey (DECaLS), the Beijing-Arizona Sky Survey (BASS), and the Mayall z-band Legacy Survey (MzLS). DECaLS, BASS and MzLS together include data obtained, respectively, at the Blanco telescope, Cerro Tololo Inter-American Observatory, NSF’s NOIRLab; the Bok telescope, Steward Observatory, University of Arizona; and the Mayall telescope, Kitt Peak National Observatory, NOIRLab. NOIRLab is operated by the Association of Universities for Research in Astronomy (AURA) under a cooperative agreement with the National Science Foundation. Pipeline processing and analyses of the data were supported by NOIRLab and the Lawrence Berkeley National Laboratory (LBNL). Legacy Surveys also uses data products from the Near-Earth Object Wide-field Infrared Survey Explorer (NEOWISE), a project of the Jet Propulsion Laboratory/California Institute of Technology, funded by the National Aeronautics and Space Administration. Legacy Surveys was supported by: the Director, Office of Science, Office of High Energy Physics of the U.S. Department of Energy; the National Energy Research Scientific Computing Center, a DOE Office of Science User Facility; the U.S. National Science Foundation, Division of Astronomical Sciences; the National Astronomical Observatories of China, the Chinese Academy of Sciences and the Chinese National Natural Science Foundation. LBNL is managed by the Regents of the University of California under contract to the U.S. Department of Energy. The complete acknowledgments can be found at \url{https://www.legacysurvey.org/acknowledgment/}.

%NERSC resources
This research used resources of the National Energy Research Scientific Computing Center, a DOE Office of Science User Facility supported by the Office of Science of the U.S. Department of Energy under Contract No. DE-AC02-05CH11231 using NERSC award HEP-ERCAP0020828.

%KiDS resources
Based in part on observations made with ESO Telescopes at the La Silla Paranal Observatory under programme IDs 177.A-3016, 177.A-3017, 177.A-3018 and 179.A-2004, and on data products produced by the KiDS consortium. The KiDS production team acknowledges support from: Deutsche Forschungsgemeinschaft, ERC, NOVA and NWO-M grants; Target; the University of Padova, and the University Federico II (Naples).


\bibliographystyle{mnras}
\bibliography{bibliography} % if your bibtex file is called example.bib

\begin{thebibliography}{}
\makeatletter
\relax
\def\mn@urlcharsother{\let\do\@makeother \do\$\do\&\do\#\do\^\do\_\do\%\do\~}
\def\mn@doi{\begingroup\mn@urlcharsother \@ifnextchar [ {\mn@doi@} {\mn@doi@[]}}
\def\mn@doi@[#1]#2{\def\@tempa{#1}\ifx\@tempa\@empty \href {http://dx.doi.org/#2} {doi:#2}\else \href {http://dx.doi.org/#2} {#1}\fi \endgroup}
\def\mn@eprint#1#2{\mn@eprint@#1:#2::\@nil}
\def\mn@eprint@arXiv#1{\href {http://arxiv.org/abs/#1} {{\tt arXiv:#1}}}
\def\mn@eprint@dblp#1{\href {http://dblp.uni-trier.de/rec/bibtex/#1.xml} {dblp:#1}}
\def\mn@eprint@#1:#2:#3:#4\@nil{\def\@tempa {#1}\def\@tempb {#2}\def\@tempc {#3}\ifx \@tempc \@empty \let \@tempc \@tempb \let \@tempb \@tempa \fi \ifx \@tempb \@empty \def\@tempb {arXiv}\fi \@ifundefined {mn@eprint@\@tempb}{\@tempb:\@tempc}{\expandafter \expandafter \csname mn@eprint@\@tempb\endcsname \expandafter{\@tempc}}}

\bibitem[\protect\citeauthoryear{{Aihara} et~al.,}{{Aihara} et~al.}{2018}]{Aihara2018}
{Aihara} H.,  et~al., 2018, \mn@doi [\pasj] {10.1093/pasj/psx066}, \href {https://ui.adsabs.harvard.edu/abs/2018PASJ...70S...4A} {70, S4}

\bibitem[\protect\citeauthoryear{Aihara et~al.,}{Aihara et~al.}{2022}]{hsc_dr3}
Aihara H.,  et~al., 2022, \mn@doi [\pasj] {10.1093/pasj/psab122}, 74, 247

\bibitem[\protect\citeauthoryear{Alsing, Peiris, Mortlock, Leja  \& Leistedt}{Alsing et~al.}{2023}]{Alsing_2023}
Alsing J.,  Peiris H.,  Mortlock D.,  Leja J.,   Leistedt B.,  2023, \mn@doi [\apjs] {10.3847/1538-4365/ac9583}, 264, 29

\bibitem[\protect\citeauthoryear{Amendola et~al.,}{Amendola et~al.}{2013}]{Euclid}
Amendola L.,  et~al., 2013, \mn@doi [Living Reviews in Relativity] {10.12942/lrr-2013-6}, 16

\bibitem[\protect\citeauthoryear{Amon et~al.,}{Amon et~al.}{2022}]{Amon_2022}
Amon A.,  et~al., 2022, \mn@doi [Phys. Rev. D] {10.1103/physrevd.105.023514}, 105

\bibitem[\protect\citeauthoryear{{Arnaboldi}, {Capaccioli}, {Mancini}, {Scaramella}, {Sedmak}  \& {Kurz}}{{Arnaboldi} et~al.}{2000}]{VST}
{Arnaboldi} M.,  {Capaccioli} M.,  {Mancini} D.,  {Scaramella} R.,  {Sedmak} G.,   {Kurz} R.,  2000, in {Bergeron} J.,  {Renzini} A.,  eds, From Extrasolar Planets to Cosmology: The VLT Opening Symposium. p.~204, \mn@doi{10.1007/10720961_27}

\bibitem[\protect\citeauthoryear{{Bailey et al.}}{{Bailey et al.}}{2023}]{redrock}
{Bailey et al.} 2023, in prep

\bibitem[\protect\citeauthoryear{{Balogh} et~al.,}{{Balogh} et~al.}{2014}]{balogh_2014}
{Balogh} M.~L.,  et~al., 2014, \mn@doi [\mnras] {10.1093/mnras/stu1332}, \href {https://ui.adsabs.harvard.edu/abs/2014MNRAS.443.2679B} {443, 2679}

\bibitem[\protect\citeauthoryear{{Brammer}, {van Dokkum}  \& {Coppi}}{{Brammer} et~al.}{2008}]{eazy}
{Brammer} G.~B.,  {van Dokkum} P.~G.,   {Coppi} P.,  2008, \mn@doi [\apj] {10.1086/591786}, \href {https://ui.adsabs.harvard.edu/abs/2008ApJ...686.1503B} {686, 1503}

\bibitem[\protect\citeauthoryear{Buchs et~al.,}{Buchs et~al.}{2019}]{Buchs_2019}
Buchs R.,  et~al., 2019, \mn@doi [\mnras] {10.1093/mnras/stz2162}, 489, 820–841

\bibitem[\protect\citeauthoryear{Casey et~al.,}{Casey et~al.}{2017}]{Casey_2017}
Casey C.~M.,  et~al., 2017, \mn@doi [\apj] {10.3847/1538-4357/aa6cb1}, 840, 101

\bibitem[\protect\citeauthoryear{Collaboration et~al.,}{Collaboration et~al.}{2009}]{lsst_science_v2}
Collaboration L.~S.,  et~al., 2009, LSST Science Book, Version 2.0 (\mn@eprint {arXiv} {0912.0201})

\bibitem[\protect\citeauthoryear{Collaboration et~al.,}{Collaboration et~al.}{2016b}]{desicollaboration2016desi_science}
Collaboration D.,  et~al., 2016b, The DESI Experiment Part I: Science,Targeting, and Survey Design (\mn@eprint {arXiv} {1611.00036})

\bibitem[\protect\citeauthoryear{Collaboration et~al.,}{Collaboration et~al.}{2016a}]{desicollaboration2016desi_inst}
Collaboration D.,  et~al., 2016a, The DESI Experiment Part II: Instrument Design (\mn@eprint {arXiv} {1611.00037})

\bibitem[\protect\citeauthoryear{{Cutri} et~al.,}{{Cutri} et~al.}{2013}]{wise_explan}
{Cutri} R.~M.,  et~al., 2013, {Explanatory Supplement to the AllWISE Data Release Products}, Explanatory Supplement to the AllWISE Data Release Products, by R. M. Cutri et al.

\bibitem[\protect\citeauthoryear{{DESI Collaboration}}{{DESI Collaboration}}{2023}]{DESI_SV_Overview}
{DESI Collaboration} 2023, \apj

\bibitem[\protect\citeauthoryear{{DESI Collaboration} et~al.,}{{DESI Collaboration} et~al.}{2022}]{desi_instrument}
{DESI Collaboration} et~al., 2022, \mn@doi [\aj] {10.3847/1538-3881/ac882b}, \href {https://ui.adsabs.harvard.edu/abs/2022AJ....164..207D} {164, 207}

\bibitem[\protect\citeauthoryear{{DESI Collaboration} et~al.,}{{DESI Collaboration} et~al.}{2023}]{desi_edr}
{DESI Collaboration} et~al., 2023, \mn@doi [arXiv e-prints] {10.48550/arXiv.2306.06308}, \href {https://ui.adsabs.harvard.edu/abs/2023arXiv230606308D} {p. arXiv:2306.06308}

\bibitem[\protect\citeauthoryear{{Dalal} et~al.,}{{Dalal} et~al.}{2023}]{hscy3_fourier}
{Dalal} R.,  et~al., 2023, \mn@doi [Phys. Rev. D] {10.1103/PhysRevD.108.123519}, 108, 123519

\bibitem[\protect\citeauthoryear{{Dark Energy Survey Collaboration} et~al.,}{{Dark Energy Survey Collaboration} et~al.}{2016}]{des_overview}
{Dark Energy Survey Collaboration} et~al., 2016, \mn@doi [\mnras] {10.1093/mnras/stw641}, \href {https://ui.adsabs.harvard.edu/abs/2016MNRAS.460.1270D} {460, 1270}

\bibitem[\protect\citeauthoryear{{Dey} et~al.,}{{Dey} et~al.}{2019}]{DESI_Legacy_Survey_DR9}
{Dey} A.,  et~al., 2019, \mn@doi [\aj] {10.3847/1538-3881/ab089d}, \href {https://ui.adsabs.harvard.edu/abs/2019AJ....157..168D} {157, 168}

\bibitem[\protect\citeauthoryear{{Driver} et~al.,}{{Driver} et~al.}{2011}]{Driver2011}
{Driver} S.~P.,  et~al., 2011, \mn@doi [\mnras] {10.1111/j.1365-2966.2010.18188.x}, \href {https://ui.adsabs.harvard.edu/abs/2011MNRAS.413..971D} {413, 971}

\bibitem[\protect\citeauthoryear{Everett et~al.,}{Everett et~al.}{2022}]{Everett_2022}
Everett S.,  et~al., 2022, \mn@doi [\apjs] {10.3847/1538-4365/ac26c1}, 258, 15

\bibitem[\protect\citeauthoryear{Flaugher \& Bebek}{Flaugher \& Bebek}{2014}]{DESI_2014}
Flaugher B.,  Bebek C.,  2014, in Ramsay S.~K.,  McLean I.~S.,   Takami H.,  eds, ~ Vol. 9147, Ground-based and Airborne Instrumentation for Astronomy V. SPIE, pp 282 -- 289, \url {https://doi.org/10.1117/12.2057105}

\bibitem[\protect\citeauthoryear{{Gebhardt} et~al.,}{{Gebhardt} et~al.}{2021}]{hetdex}
{Gebhardt} K.,  et~al., 2021, \mn@doi [\apj] {10.3847/1538-4357/ac2e03}, \href {https://ui.adsabs.harvard.edu/abs/2021ApJ...923..217G} {923, 217}

\bibitem[\protect\citeauthoryear{Gruen \& Brimioulle}{Gruen \& Brimioulle}{2017}]{gruen_2017}
Gruen D.,  Brimioulle F.,  2017, \mn@doi [\mnras] {10.1093/mnras/stx471}, 468, 769–782

\bibitem[\protect\citeauthoryear{Gruen \& McCullough}{Gruen \& McCullough}{2023}]{4c3r2}
Gruen D.,  McCullough J.,  2023, The Messenger, 190

\bibitem[\protect\citeauthoryear{Guy et~al.,}{Guy et~al.}{2023}]{desi_spec_pipeline}
Guy J.,  et~al., 2023, \mn@doi [\aj] {10.3847/1538-3881/acb212}, 165, 144

\bibitem[\protect\citeauthoryear{{Hahn} et~al.,}{{Hahn} et~al.}{2023}]{BGS_paper}
{Hahn} C.,  et~al., 2023, \mn@doi [\aj] {10.3847/1538-3881/accff8}, \href {https://ui.adsabs.harvard.edu/abs/2023AJ....165..253H} {165, 253}

\bibitem[\protect\citeauthoryear{Hartley et~al.,}{Hartley et~al.}{2020}]{Hartley_2020}
Hartley W.~G.,  et~al., 2020, \mn@doi [\mnras] {10.1093/mnras/staa1812}, 496, 4769

\bibitem[\protect\citeauthoryear{Hasinger et~al.,}{Hasinger et~al.}{2018}]{Hasinger_2018}
Hasinger G.,  et~al., 2018, \mn@doi [\apj] {10.3847/1538-4357/aabacf}, 858, 77

\bibitem[\protect\citeauthoryear{Hildebrandt et~al.,}{Hildebrandt et~al.}{2020}]{Hildebrandt_2020}
Hildebrandt H.,  et~al., 2020, \mn@doi [\aap] {10.1051/0004-6361/201834878}, 633, A69

\bibitem[\protect\citeauthoryear{Huang et~al.,}{Huang et~al.}{2017}]{Huang_2017}
Huang S.,  et~al., 2017, \mn@doi [\pasj] {10.1093/pasj/psx126}, 70

\bibitem[\protect\citeauthoryear{{Ivezi{\'c}} et~al.,}{{Ivezi{\'c}} et~al.}{2019}]{Rubin}
{Ivezi{\'c}} {\v Z}.,  et~al., 2019, \mn@doi [\apj] {10.3847/1538-4357/ab042c}, \href {http://adsabs.harvard.edu/abs/2019ApJ...873..111I} {873, 111}

\bibitem[\protect\citeauthoryear{Kannawadi et~al.,}{Kannawadi et~al.}{2019}]{Kannawadi_2019}
Kannawadi A.,  et~al., 2019, \mn@doi [\aap] {10.1051/0004-6361/201834819}, 624, A92

\bibitem[\protect\citeauthoryear{{Kartaltepe} et~al.,}{{Kartaltepe} et~al.}{2010}]{kartaltepe_2010}
{Kartaltepe} J.~S.,  et~al., 2010, \mn@doi [\apj] {10.1088/0004-637X/709/2/572}, \href {https://ui.adsabs.harvard.edu/abs/2010ApJ...709..572K} {709, 572}

\bibitem[\protect\citeauthoryear{Kohonen}{Kohonen}{2004}]{som}
Kohonen T.,  2004, Biological Cybernetics, 43, 59

\bibitem[\protect\citeauthoryear{{Kriek} et~al.,}{{Kriek} et~al.}{2015}]{kriek_2015}
{Kriek} M.,  et~al., 2015, \mn@doi [\apjs] {10.1088/0067-0049/218/2/15}, \href {https://ui.adsabs.harvard.edu/abs/2015ApJS..218...15K} {218, 15}

\bibitem[\protect\citeauthoryear{{Kuijken, K.} et~al.,}{{Kuijken, K.} et~al.}{2019}]{KiDS_DR4}
{Kuijken, K.} et~al., 2019, \mn@doi [\aap] {10.1051/0004-6361/201834918}, 625, A2

\bibitem[\protect\citeauthoryear{{Kuijken} et~al.,}{{Kuijken} et~al.}{2015}]{kids_vst}
{Kuijken} K.,  et~al., 2015, \mn@doi [\mnras] {10.1093/mnras/stv2140}, \href {https://ui.adsabs.harvard.edu/abs/2015MNRAS.454.3500K} {454, 3500}

\bibitem[\protect\citeauthoryear{{LSST Dark Energy Science Collaboration} et~al.,}{{LSST Dark Energy Science Collaboration} et~al.}{2018}]{lsstreqs2021}
{LSST Dark Energy Science Collaboration} et~al., 2018, The LSST Dark Energy Science Collaboration (DESC) Science Requirements Document, \mn@doi{10.48550/ARXIV.1809.01669}, \url {https://arxiv.org/abs/1809.01669}

\bibitem[\protect\citeauthoryear{{Laigle} et~al.,}{{Laigle} et~al.}{2016}]{Laigle2016}
{Laigle} C.,  et~al., 2016, \mn@doi [\apjs] {10.3847/0067-0049/224/2/24}, \href {https://ui.adsabs.harvard.edu/abs/2016ApJS..224...24L} {224, 24}

\bibitem[\protect\citeauthoryear{Lan et~al.,}{Lan et~al.}{2023}]{VI_paper}
Lan T.-W.,  et~al., 2023, \mn@doi [\apj] {10.3847/1538-4357/aca5fa}, 943, 68

\bibitem[\protect\citeauthoryear{{Le F{\`e}vre} et~al.,}{{Le F{\`e}vre} et~al.}{2015}]{LeFevre}
{Le F{\`e}vre} O.,  et~al., 2015, \mn@doi [\aap] {10.1051/0004-6361/201423829}, \href {https://ui.adsabs.harvard.edu/abs/2015A&A...576A..79L} {576, A79}

\bibitem[\protect\citeauthoryear{{Li} et~al.,}{{Li} et~al.}{2023a}]{hscy3_real}
{Li} X.,  et~al., 2023a, \mn@doi [\prd] {10.1103/PhysRevD.108.123518}, \href {https://ui.adsabs.harvard.edu/abs/2023PhRvD.108l3518L} {108, 123518}

\bibitem[\protect\citeauthoryear{Li et~al.,}{Li et~al.}{2023b}]{li2022}
Li S.-S.,  et~al., 2023b, \mn@doi [\aap] {10.1051/0004-6361/202245210}, 670, A100

\bibitem[\protect\citeauthoryear{{Lilly} et~al.,}{{Lilly} et~al.}{2007}]{Lilly2007}
{Lilly} S.~J.,  et~al., 2007, \mn@doi [\apjs] {10.1086/516589}, \href {https://ui.adsabs.harvard.edu/abs/2007ApJS..172...70L} {172, 70}

\bibitem[\protect\citeauthoryear{Masters et~al.,}{Masters et~al.}{2015}]{Masters_2015}
Masters D.,  et~al., 2015, \mn@doi [\apj] {10.1088/0004-637x/813/1/53}, 813, 53

\bibitem[\protect\citeauthoryear{Masters, Stern, Cohen, Capak, Rhodes, Castander  \& Paltani}{Masters et~al.}{2017}]{Masters_2017}
Masters D.~C.,  Stern D.~K.,  Cohen J.~G.,  Capak P.~L.,  Rhodes J.~D.,  Castander F.~J.,   Paltani S.,  2017, \mn@doi [\apj] {10.3847/1538-4357/aa6f08}, 841, 111

\bibitem[\protect\citeauthoryear{Masters et~al.,}{Masters et~al.}{2019}]{Masters_2019}
Masters D.~C.,  et~al., 2019, \mn@doi [\apj] {10.3847/1538-4357/ab184d}, 877, 81

\bibitem[\protect\citeauthoryear{Myers et~al.,}{Myers et~al.}{2023}]{targsel_paper}
Myers A.~D.,  et~al., 2023, \mn@doi [\aj] {10.3847/1538-3881/aca5f9}, 165, 50

\bibitem[\protect\citeauthoryear{Myles et~al.,}{Myles et~al.}{2021}]{Myles_2021}
Myles J.,  et~al., 2021, \mn@doi [\mnras] {10.1093/mnras/stab1515}, 505, 4249–4277

\bibitem[\protect\citeauthoryear{Newman \& Gruen}{Newman \& Gruen}{2022}]{GruenNewman}
Newman J.~A.,  Gruen D.,  2022, \mn@doi [ARA\&A] {10.1146/annurev-astro-032122-014611}, 60, 363

\bibitem[\protect\citeauthoryear{{Oke} \& {Gunn}}{{Oke} \& {Gunn}}{1983}]{AB}
{Oke} J.~B.,  {Gunn} J.~E.,  1983, \mn@doi [\apj] {10.1086/160817}, \href {https://ui.adsabs.harvard.edu/abs/1983ApJ...266..713O} {266, 713}

\bibitem[\protect\citeauthoryear{Oke et~al.,}{Oke et~al.}{1995}]{keck_LRS}
Oke J.~B.,  et~al., 1995, \mn@doi [PASP] {10.1086/133562}, 107, 375

\bibitem[\protect\citeauthoryear{Raichoor et~al.,}{Raichoor et~al.}{2023}]{ELG_paper}
Raichoor A.,  et~al., 2023, \mn@doi [\apj] {10.3847/1538-3881/acb213}, 165, 126

\bibitem[\protect\citeauthoryear{Rau et~al.,}{Rau et~al.}{2023}]{Rau23}
Rau M.~M.,  et~al., 2023, \mn@doi [\mnras] {10.1093/mnras/stad1962}, 524, 5109

\bibitem[\protect\citeauthoryear{Rykoff et~al.,}{Rykoff et~al.}{2014}]{Rykoff_2014}
Rykoff E.~S.,  et~al., 2014, \mn@doi [\apj] {10.1088/0004-637X/785/2/104}, 785, 104

\bibitem[\protect\citeauthoryear{Rykoff et~al.,}{Rykoff et~al.}{2016}]{Rykoff_2016}
Rykoff E.~S.,  et~al., 2016, \mn@doi [\apjs] {10.3847/0067-0049/224/1/1}, 224, 1

\bibitem[\protect\citeauthoryear{Saglia et~al.,}{Saglia et~al.}{2022}]{euclidprep}
Saglia R.,  et~al., 2022, \mn@doi [\aap] {10.1051/0004-6361/202243604}, 664, A196

\bibitem[\protect\citeauthoryear{{Salvato}, {Ilbert}  \& {Hoyle}}{{Salvato} et~al.}{2019}]{salvato2018flavours}
{Salvato} M.,  {Ilbert} O.,   {Hoyle} B.,  2019, \mn@doi [Nature Astronomy] {10.1038/s41550-018-0478-0}, \href {https://ui.adsabs.harvard.edu/abs/2019NatAs...3..212S} {3, 212}

\bibitem[\protect\citeauthoryear{Schlegel et~al.,}{Schlegel et~al.}{2022}]{snowmass_desiII}
Schlegel D.~J.,  et~al., 2022, A Spectroscopic Road Map for Cosmic Frontier: DESI, DESI-II, Stage-5 (\mn@eprint {arXiv} {2209.03585})

\bibitem[\protect\citeauthoryear{Sevilla-Noarbe et~al.,}{Sevilla-Noarbe et~al.}{2021}]{desy3_photom}
Sevilla-Noarbe I.,  et~al., 2021, \mn@doi [\apjs] {10.3847/1538-4365/abeb66}, 254, 24

\bibitem[\protect\citeauthoryear{{Silverman} et~al.,}{{Silverman} et~al.}{2015}]{silverman_2015}
{Silverman} J.~D.,  et~al., 2015, \mn@doi [\apjs] {10.1088/0067-0049/220/1/12}, \href {https://ui.adsabs.harvard.edu/abs/2015ApJS..220...12S} {220, 12}

\bibitem[\protect\citeauthoryear{{Stanford} et~al.,}{{Stanford} et~al.}{2021}]{c3r2_dr3}
{Stanford} S.~A.,  et~al., 2021, \mn@doi [\apjs] {10.3847/1538-4365/ac0833}, \href {https://ui.adsabs.harvard.edu/abs/2021ApJS..256....9S} {256, 9}

\bibitem[\protect\citeauthoryear{{Trump} et~al.,}{{Trump} et~al.}{2007}]{trump_2007}
{Trump} J.~R.,  et~al., 2007, \mn@doi [\apjs] {10.1086/516578}, \href {https://ui.adsabs.harvard.edu/abs/2007ApJS..172..383T} {172, 383}

\bibitem[\protect\citeauthoryear{Wang et~al.,}{Wang et~al.}{2023}]{Wang_2023}
Wang B.,  et~al., 2023, \mn@doi [APJL] {10.3847/2041-8213/acba99}, 944, L58

\bibitem[\protect\citeauthoryear{Wright et~al.,}{Wright et~al.}{2010}]{Wright_2010}
Wright E.~L.,  et~al., 2010, \mn@doi [\aj] {10.1088/0004-6256/140/6/1868}, 140, 1868

\bibitem[\protect\citeauthoryear{Wright et~al.,}{Wright et~al.}{2019}]{Wright_2019}
Wright A.~H.,  et~al., 2019, \mn@doi [\aap] {10.1051/0004-6361/201834879}, 632, A34

\bibitem[\protect\citeauthoryear{Wright, Hildebrandt, van~den Busch, Heymans, Joachimi, Kannawadi  \& Kuijken}{Wright et~al.}{2020}]{Wright_2020}
Wright A.~H.,  Hildebrandt H.,  van~den Busch J.~L.,  Heymans C.,  Joachimi B.,  Kannawadi A.,   Kuijken K.,  2020, \mn@doi [\aap] {10.1051/0004-6361/202038389}, 640, L14

\bibitem[\protect\citeauthoryear{Zhou et~al.,}{Zhou et~al.}{2023}]{LRG_paper}
Zhou R.,  et~al., 2023, \mn@doi [\aj] {10.3847/1538-3881/aca5fb}, 165, 58

\bibitem[\protect\citeauthoryear{Zou et~al.,}{Zou et~al.}{2017}]{Zou_2017}
Zou H.,  et~al., 2017, \mn@doi [\pasp] {10.1088/1538-3873/aa65ba}, 129, 064101

\bibitem[\protect\citeauthoryear{de Jong et~al.,}{de~Jong et~al.}{2017}]{kidsDR3_de_Jong_2017}
de Jong J. T.~A.,  et~al., 2017, \mn@doi [A\&A] {10.1051/0004-6361/201730747}, 604, A134

\bibitem[\protect\citeauthoryear{de Jong et~al.,}{de~Jong et~al.}{2019}]{4most}
de Jong R.,  et~al., 2019, The Messenger, 175

\bibitem[\protect\citeauthoryear{van~den Busch et~al.,}{van~den Busch et~al.}{2022}]{van_den_Busch_2022}
van~den Busch J.~L.,  et~al., 2022, \mn@doi [A\&A] {10.1051/0004-6361/202142083}, 664, A170

\makeatother
\end{thebibliography}

\section*{Affiliations}
\noindent
{\it
$^{1}$Kavli Institute for Particle Astrophysics and Cosmology, Department of Physics, Stanford University, Stanford, CA, USA\\
$^{2}$SLAC National Accelerator Laboratory, Menlo Park, CA, USA\\
$^{3}$Universitäts-Sternwarte, Fakultät für Physik, Ludwig-Maximilians-Universität München,
Scheinerstraße 1, 81679 München, Germany\\
$^{4}$Excellence Cluster ORIGINS, Boltzmannstrasse 2, D-85748 Garching, Germany\\
$^{5}$Institute of Astronomy, University of Cambridge, Madingley Road, Cambridge CB3 0HA, UK \\%alex
$^{6}$Kavli Institute for Cosmology, University of Cambridge, Madingley Road, Cambridge CB3 0HA, UK \\%alex
$^{7}$California institute of Technology, 1200 E California Blvd, Pasadena, CA 91125, USA \\%masters
$^{8}$Lawrence Berkeley National Laboratory, 1 Cyclotron Road, Berkeley, CA 94720, USA \\%schlegel, derose, anand, wilson
% $^{8}$Centre for Astrophysics \& Supercomputing, Swinburne University of Technology, P.O. Box 218, Hawthorn, VIC 3122, Australia \\%blake
$^{9}$Institute of Cosmology \& Gravitation, University of Portsmouth, Dennis Sciama Building, Portsmouth, PO1 3FX, UK \\%canning
$^{10}$Institut d’Estudis Espacials de Catalunya (IEEC), 08034 Barcelona, Spain \\%castander
$^{11}$ Institute of Space Sciences (ICE, CSIC), Campus UAB, Carrer de Can Magrans, 08193 Barcelona, Spain\\ %castander
$^{12}$Institució Catalana de Recerca i Estudis Avançats,E-08010 Barcelona, Spain\\ %miquel
$^{13}$Institut de Física d’Altes Energies (IFAE), The Barcelona Institute of Science and Technology, Campus UAB,08193 Bellaterra (Barcelona) Spain\\ %miquel
$^{14}$Department of Physics and Astronomy and PITT PACC, University of Pittsburgh, Pittsburgh, PA 15260, USA\\ %newman
$^{15}$Brookhaven National Laboratory, Upton, NY 11973\\ %slosar
$^{16}$University of Toronto, Dunlap Institute for Astronomy \& Astrophysics, Toronto, M5S 3H4, Canada \\%speagle
$^{17}$Institute for Computational Cosmology, Department of Physics, Durham University, South Road, Durham DH1 3LE, UK \\%wilson
$^{18}$Physics Dept., Boston University, 590 Commonwealth Avenue, Boston, MA 02215, USA\\
$^{19}$Department of Physics \& Astronomy, University College London, Gower Street, London, WC1E 6BT, UK\\
$^{20}$Department of Physics and Astronomy, The University of Utah, 115 South 1400 East, Salt Lake City, UT 84112, USA \\
$^{21}$Instituto de F\'{\i}sica, Universidad Nacional Aut\'{o}noma de M\'{e}xico,  Cd. de M\'{e}xico  C.P. 04510,  M\'{e}xico\\
$^{22}$Departamento de F\'isica, Universidad de los Andes, Cra. 1 No. 18A-10, Edificio Ip, CP 111711, Bogot\'a, Colombia\\
$^{23}$Department of Physics, Southern Methodist University, 3215 Daniel Avenue, Dallas, TX 75275, USA \\
$^{24}$Sorbonne Universit\'{e}, CNRS/IN2P3, Laboratoire de Physique Nucl\'{e}aire et de Hautes Energies (LPNHE), FR-75005 Paris, France \\
$^{25}$Center for Cosmology and AstroParticle Physics, The Ohio State University, 191 West Woodruff Avenue, Columbus, OH 43210, USA\\
$^{26}$NSF's NOIRLab, 950 N. Cherry Ave., Tucson, AZ 85719, USA \\
$^{27}$Department of Physics and Astronomy, Siena College, 515 Loudon Road, Loudonville, NY 12211, USA\\
$^{28}$National Astronomical Observatories, Chinese Academy of Sciences, A20 Datun Rd., Chaoyang District, Beijing, 100012, P.R. China\\
$^{29}$Department of Physics and Astronomy, University of Waterloo, 200 University Ave W, Waterloo, ON N2L 3G1, Canada\\
$^{30}$Space Sciences Laboratory, University of California, Berkeley, 7 Gauss Way, Berkeley, CA  94720, USA\\
$^{31}$Instituto de Astrof\'{i}sica de Andaluc\'{i}a (CSIC), Glorieta de la Astronom\'{i}a, s/n, E-18008 Granada, Spain\\
$^{32}$Department of Physics, Kansas State University, 116 Cardwell Hall, Manhattan, KS 66506, USA\\
$^{33}$Department of Physics and Astronomy, Sejong University, Seoul, 143-747, Korea\\
$^{34}$CIEMAT, Avenida Complutense 40, E-28040 Madrid, Spain\\
$^{35}$Department of Physics \& Astronomy, Ohio University, Athens, OH 45701, USA\\
$^{36}$University of Michigan, Ann Arbor, MI 48109, USA\\
$^{37}$National Astronomical Observatories, Chinese Academy of Sciences, A20 Datun Rd., Chaoyang District, Beijing, 100012, P.R. China\\
}

%%%%%%%%%%%%%%%%%%%%%%%%%%%%%%%%%%%%%%%%%%%%%%%%%%

%%%%%%%%%%%%%%%%% APPENDICES %%%%%%%%%%%%%%%%%%%%%
%\clearpage
\appendix
\section{The SOM \& color-space}
\label{app:som}
\begin{figure*}
    \centering
    \includegraphics[width=2\columnwidth]{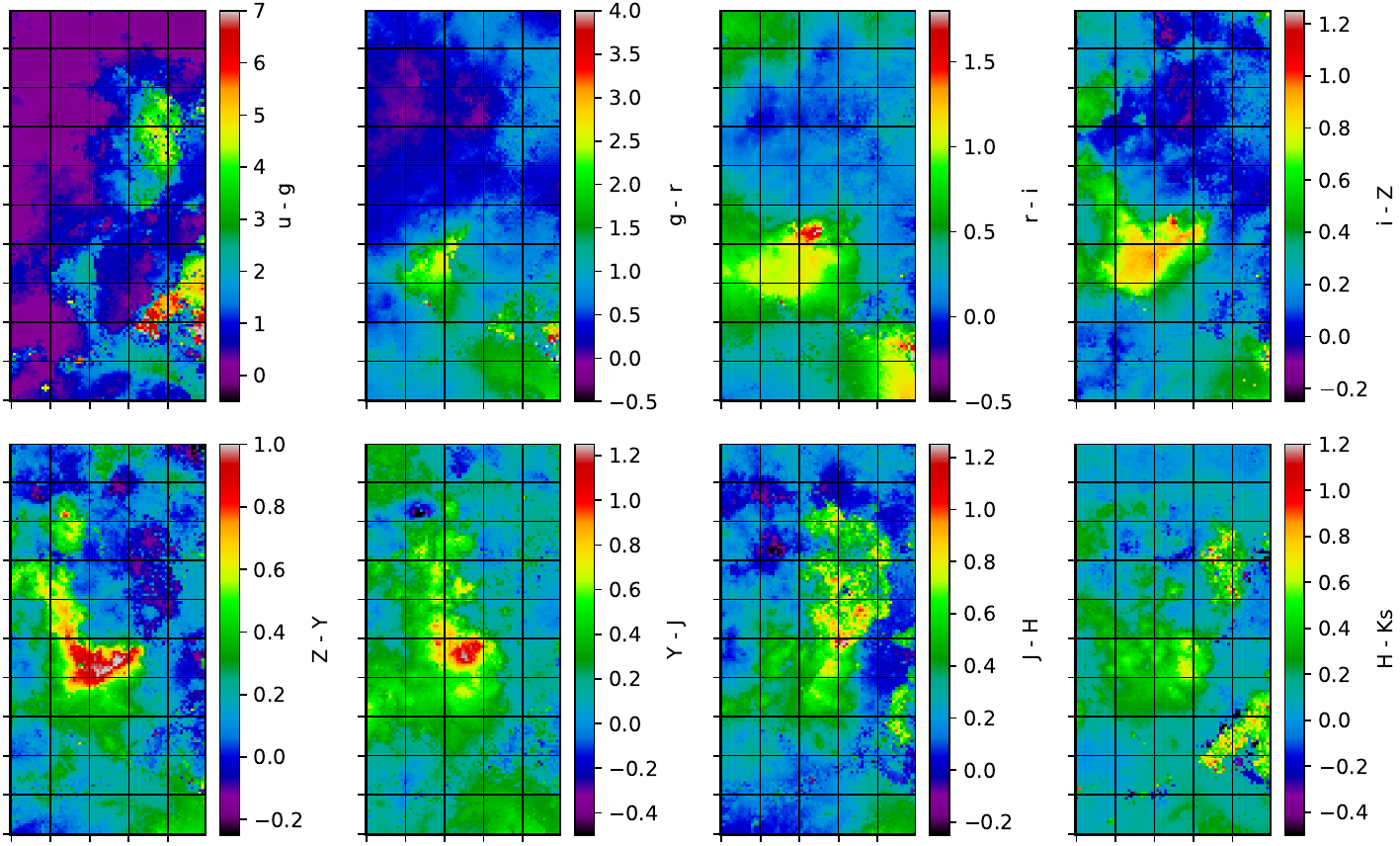}
     \caption{Depiction of the \citealt{Masters_2017} SOM corrected to KiDS-VIKING $ugriZYJHK_s$ colors, as used for this analysis.}
    \label{fig:kids_som_colors}
\end{figure*}

The SOM described in Sec. \ref{sec:verification} is fully described in Fig. \ref{fig:kids_som_colors} in the KiDS-VIKING \textit{ugriZYJHKs} colors, transformed from the original \citealt{Masters_2017} SOM using COSMOS narrowband best fit SEDs re-evaluated in the KiDS-VIKING filters.

\begin{figure*}
    \centering
    \includegraphics[width=2\columnwidth]{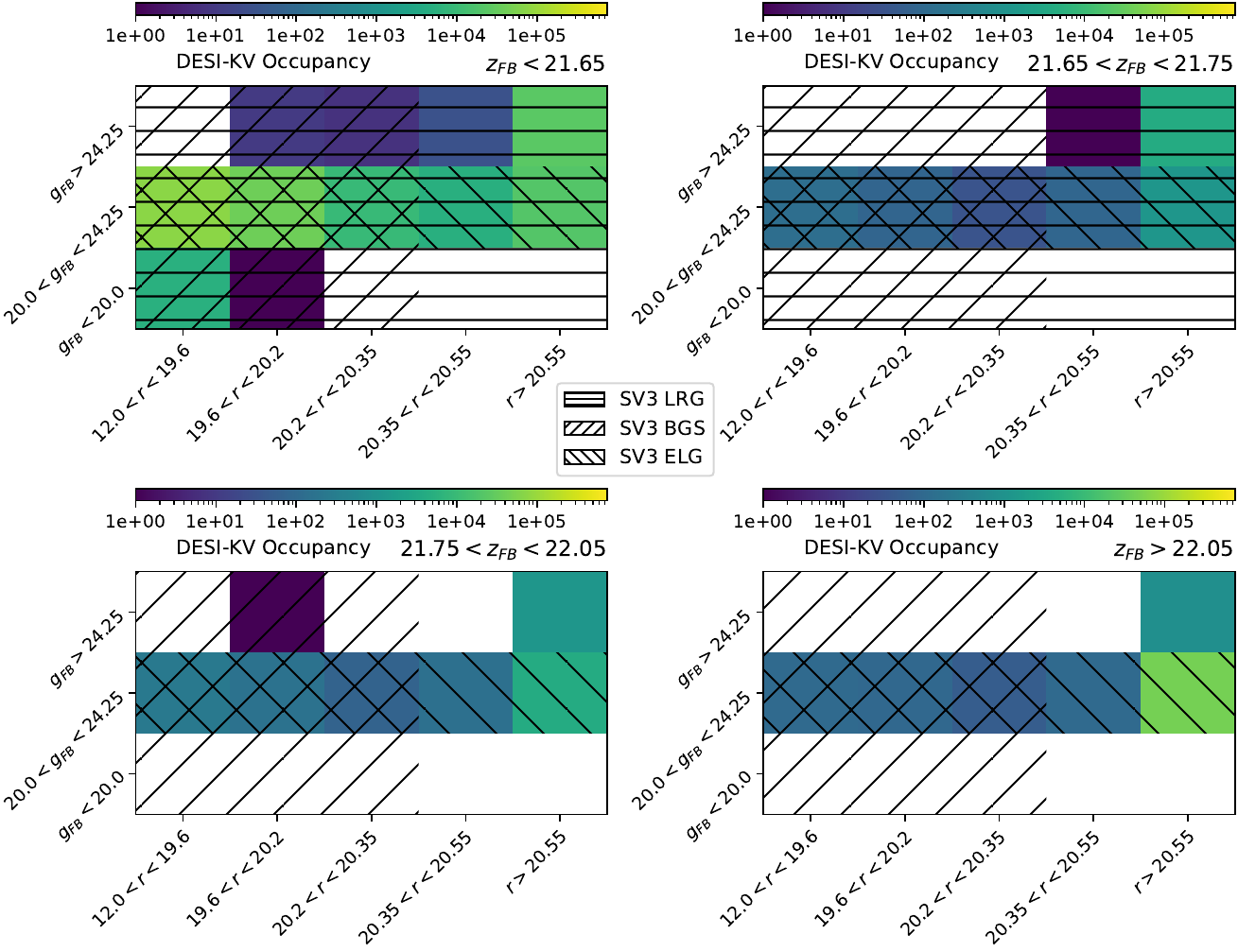}
     \caption{Depiction of the occupancy of the entire DESI-KV sample in the \textit{grz} magnitude bins (3 x 6 x 4) used to produce the weight scheme in Sec. \ref{sec:weights} that renders cell-level redshift distributions more representative.}
    \label{fig:magbins_all}
\end{figure*}

Additionally, the complete set of magnitude bins chosen for the weighting schema in Sec. \ref{sec:weights} are depicted in Fig. \ref{fig:magbins_all}, with the bin edges tuned to encapsulate sample selection boundaries in each band. The hatch patterns specify the SV3 selection boundaries for each sample, while the main and SV1 selections have mild differences. The occupancy of these magnitude color bins heavily depends on the quantity of each main survey target.

\section{SOM-C3R2 Agreement}
\label{app:som_verif}
The $p(z|c)$ distributions in a given cell may differ in the original C3R2 SOM \citep{Masters_2017} to that from the map used for this work. Combining results from differing surveys should come with the following caveats, with systematic factors contributing to changes in redshift distributions: 
\begin{figure}
	\includegraphics[width=\columnwidth]{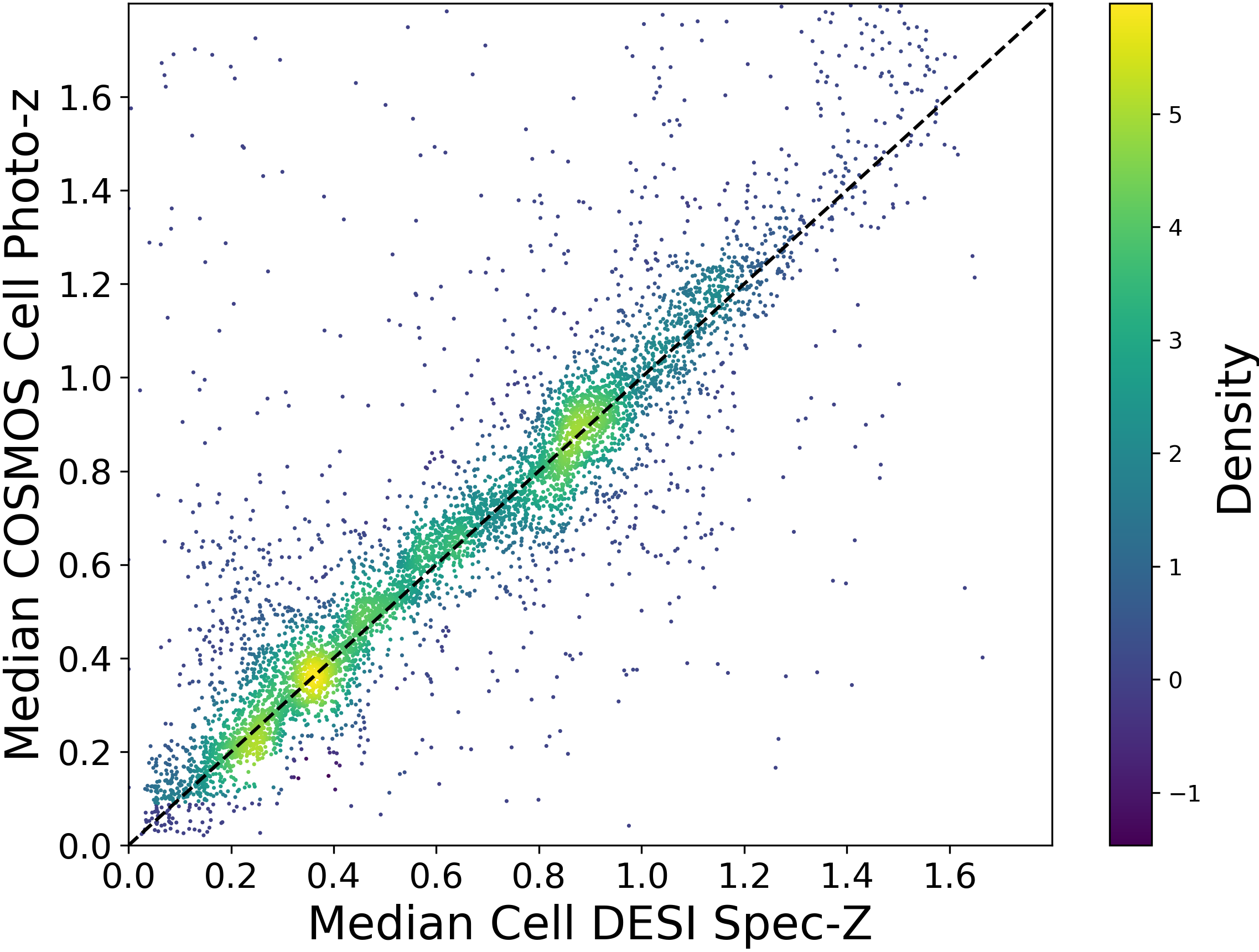}
    \caption{Median DC3R2 redshifts of a SOM cell in the transformed KV mapping compared to the COSMOS photometric redshift of the same SOM cell prior to the transformation. This reveals the effect of both the SOM color transformation as well as the differences in sample depth. Broadly speaking, there is good recovery between redshifts for objects of the same colors.}
    \label{fig:spec_v_phot}
\end{figure}
\begin{itemize}
    \item \textbf{Cosmic variance - }{The original C3R2 survey observes a different limited volumes of the Universe compared to the the overlap region of DESI and KiDS-VIKING used for DC3R2, with galaxy samples of limited number. This difference in cosmic variance can influence shifts in the redshift distributions seen in each cell. \rev{If each cell is well described as $\delta(z)$, this impact of altered abundances is minimal on the calibrated color-redshift relation. This paper has found that for the photometry used, the redshift distributions are comparably broader, in which case altered abundances may overrepresent certain contributors to the overall cell redshift. A small sample is also liable to miss cell contributors that are rare and may not accurately measure the tails of these SOM cell distributions.} However, the DESI/KV overlapping footprint is sufficiently large that cosmic variance will be a minor concern.}
    
    \item \textbf{Photometric noise - } {KV photometry has a higher noise level than the COSMOS catalog that used to train the original the map. The effect of this a scattering of galaxies between neighboring cells, effectively lowering the resolution of the map. This effect would influence  measurable quantities like $dz/dm$, which is discuss further in Section~\ref{sec:scatter}, as this noise is asymmetric between cells.}
    
    \item \textbf{The SOM color transformation - } {As the filters covered by KiDS-VIKING are different from those used in \cite{Masters_2017}, a best fit SED for each cell was defined based on template fits to COSMOS \rev{narrow-band galaxies in \cite{Laigle2016} and} evaluated in the relevant KiDS-VIKING bands to transform the color of that cell. \rev{The median of all COSMOS galaxies in the original SOM produced the best fit SEDs for this evaluation.}For a sample with overlap in COSMOS/KiDS, we ought to show that the same galaxies reliably get assigned to the same cell given sufficiently low photometric uncertainty. Before a joint analysis of the samples assigned to cells based on their original COSMOS and based on KV photometry could be made, which we do not attempt here, it would first have to be shown that the color transformation does not introduce e.g. spurious magnitude dependence of redshift within color cells. We can see a broad recovery of COSMOS cell photo-z against the DESI spectroscopic median redshift per cell in Fig. \ref{fig:spec_v_phot}, which indicates that this transformation produces a color map with very similar galaxies on a cell-by-cell basis. However, cell boundaries in color space could have minor distortions. \rev{Outliers at high $z$ are likely to result from spectroscopic incompleteness in $z \geq 1.6$. Photometric scatter is likely the largest contributor to disagreement at lower redshifts, which we elaborate on the consequences of more fully in Sec. \ref{sec:magdepend} and care to characterize for future surveys.}}
    \item \textbf{Differing redshift fitting algorithms in spectral modeling - } {Robustness tests on DESI's spectral modeling pipeline, \textsc{redrock} \citep{redrock}, are expanded upon in \cite{desi_spec_pipeline, VI_paper}. Additionally, for a small overlap of objects in common fields, we can directly compare the \textsc{redrock} modelled redshift to those found by other surveys to test their reliability.}
    \item \textbf{Spectroscopic selection effects - } {Quality flags chief among them, we can cross check on other spectral surveys to inform our redshift quality cuts that limit outliers. Perhaps even more importantly we can look at our relevant findings as a function of \textsc{redrock}'s $\Delta \chi^2$ to see if this selection affects quantities like \textit{dz/dm} or shifts our understanding of the $P(z|c)$. Certain galaxy types are more conducive to confident redshifts. Past studies have found that if color cuts are implemented on spectroscopic galaxies that are not available to the weak lensing surveys, redshift calibration errors can be as large as $\Delta z\approx 0.04$ \cite{Hartley_2020}. This analysis mitigates this effect by removal of detections that do not have sufficiently high completeness in a given cell which accounts for these sample specific color cuts (see Section \ref{sec:weights}).
    \item \textbf{Cell assignment algorithm} {The metric that is minimized for each galaxy to be assigned a cell can dramatically impact the distribution across the map. We attempt to imitate the cell assignment procedure from \cite{Masters_2015}, which minimizes contributions from flux measurements with larger error and from drop-outs, and when running the assignment on the original COSMOS catalog recover the identical assignment for more than 99\% match.}}
\end{itemize}

\section{Dedicated Tile Target Optimization}
\label{sec:priority}

The dedicated tiles comprise of two equatorial pointings at 217.5 degrees and 221.0 degrees RA, with a bright (30 minutes, \revtwo{$z_{\rm fiber} < 21.5$}) and faint (3x30 minutes, \revtwo{$z_{\rm fiber} < 22.10$}) set of exposures each. A mix of bright and faint targets were selected according to the procedure described here.\par

The aim of target choice is to optimize our measurement of $dz/dm$ within a cell, and measure it for as many color-cells as possible. We devise a metric to select targets that best do this between the bright and faint selection criteria. A function that describes the energy of our targeting procedure can be written below that optimizes our fiber choice when minimized. We chose the following energy function, $E$:
\begin{equation}
    E = \frac{1}{N_{\rm{cells}}} Var\left(\left<\frac{dz}{dm}\right>\right) = \frac{1}{N_{\rm{cells}}(>1\ {\rm target})\sum_{i\in N} 1/\sigma_i^2}.
\end{equation}
Where the variance of the measurement of the slope, $\sigma^2$ of a given cell, $i$ is proportional to
\begin{equation}
    \sigma_i^2 \propto \frac{1}{\sum_{j\in \rm{cell}\ i} \left(Z_j - \bar{Z}\right)^2}
\end{equation}
That prioritizes having a large span in $\Delta Z$ per cell, as well as maximal coverage in the color space, $N_{\rm{cell}}$, with multiple targets per cell. A simple simulated annealing algorithm was implemented to minimize this function, by varying choice of target on a fiber-by-fiber basis. For certain fibers, only a single target will be visible in both the bright and faint exposure. The optimization becomes useful when multiple targets are visible to each fiber. 

\begin{comment}
\begin{figure}
	\includegraphics[width=\columnwidth]{draft_figures/optimized_targets.png}
    \caption{Z-band fiber magnitudes for the target optimization procedure.\note{Maybe useful to use absolute distributions rather than normalized here}}
    \label{fig:annealing}
\end{figure}
\end{comment}

\section{Scaling of Exposure Time Map}
\label{sec:exptime}

The exposure times in Fig.~\ref{fig:exptime} and corresponding templates were drawn from a diverse sample with varying observation conditions and fluxes across a given color. The rationale for the reported Z = 21.0 magnitude, is that for a point source in a sky limited background, these exposure times might be scaled accordingly to estimate future spectral survey needs, e.g. 4MOST, \citep{4most}. For a sky limited background,
\begin{equation}
    SNR \propto \sqrt{t_{\mathrm{exp}}}
\end{equation}
where $t_{exp}$ is the total exposure time for a given object. In order to scale existing exposure time measurements in this regime to those that would be necessary for a Z = 21.0 object, we use the following prescription,
\begin{equation}
    t_{\mathrm{exp,sc}} = 10^{0.8(21.0 - Z_{\mathrm{o}})}t_{\mathrm{exp,o}}
\end{equation}
where $t_{\mathrm{exp,sc}},\ t_{\mathrm{exp,o}}$ are the scaled and original exposure times respectively, and $Z_{\mathrm{o}}$ is the Z magnitude of the object being scaled. This follows from the sky limited case for a point source and ought to be treated as a rough estimate of the exposure times necessary. Additionally, it is useful to develop a scaling for the quality of the redshift. As the \textsc{redrock} quantity of DELTACHI2, $\Delta \chi^2$ is our primary quality cut for redshift completeness, and the most extreme requirement for any galaxy is $\Delta \chi^2 > 40$. We can add another factor of $\frac{40}{{\Delta \chi^2}_{\mathrm{o}}}$ to account for this, with some scatter. Thus, Fig. \ref{fig:exptime} also includes projections from low-confidence redshift targets, where our overall analysis does not. These are targets that have been observed at low signal-to-noise, or lack helpful spectral features in the available wavelength range. There may thus be model uncertainty between several potential redshifts or spectral templates, rendering realistic projections for exposure times in cells that are otherwise incomplete.

\par Our scaling relationship follows as an approximation,
\begin{equation}
    t_{\mathrm{exp,sc}} \approx \frac{40}{{\Delta \chi^2}_{\mathrm{o}}}10^{0.8(21.0 - Z_{\mathrm{o}})}t_{\mathrm{exp,o}}
\end{equation}

\section{$dz/dm$ measurement}
\label{app:dzdm_slope}
\begin{figure}
    \centering
    \includegraphics[width=\columnwidth]{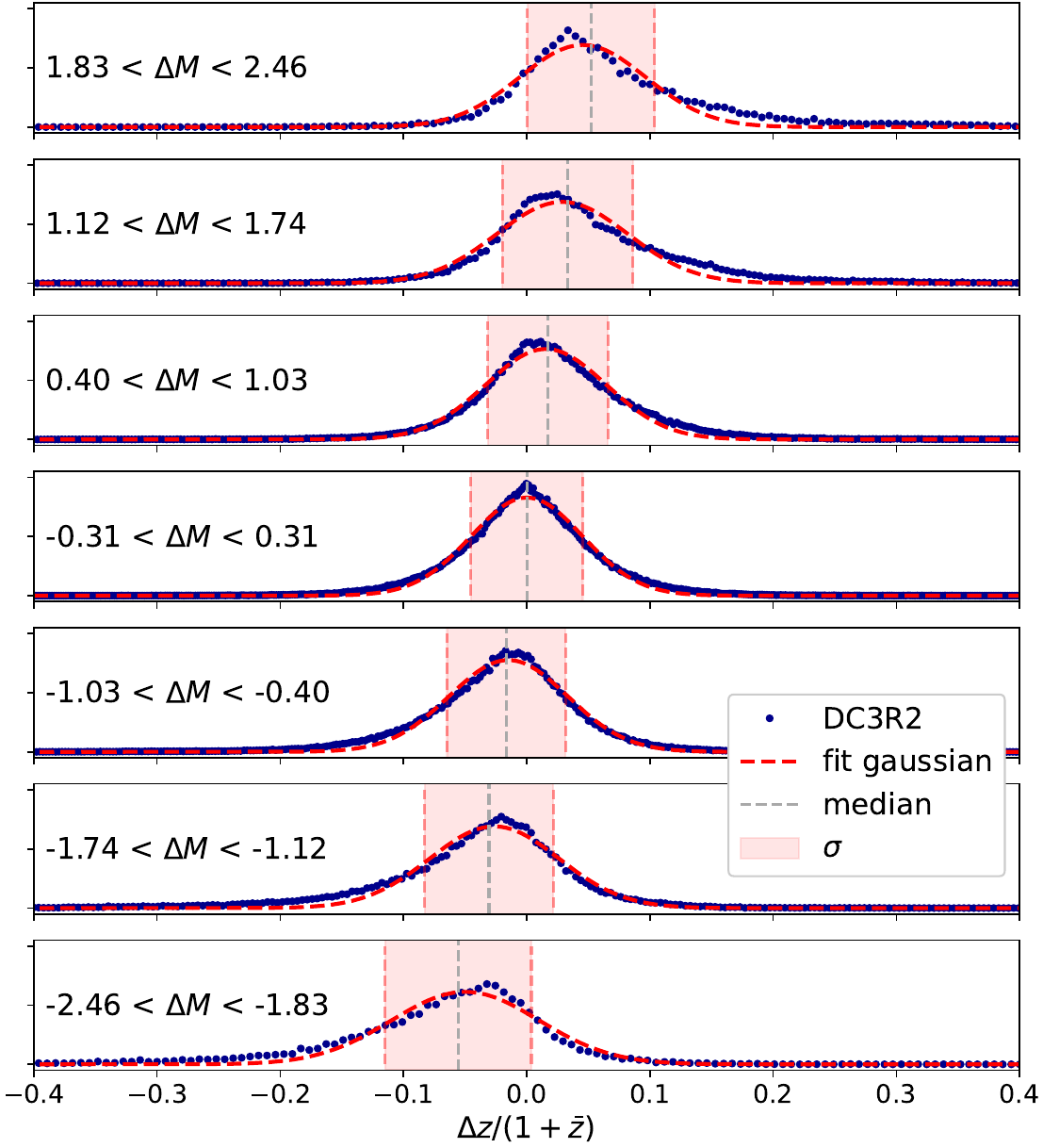}
    \caption{Example set of histograms (blue) for each magnitude bin, with median (gray) and Gaussian fits (red) to measure $dz/dm$. While there is some skew in the most extreme magnitude bins, the Gaussian is a good approximation, for which the fit $\sigma$ describes the width.}
    \label{fig:appdzdm}
\end{figure}

The slopes measured in Sec. \ref{sec:magdepend} for $dz/dm$ were done in such a way to be comparable to the procedure for the original measurement in \citealt{Masters_2017}. The data points in Fig. \ref{fig:dzdmbig} are drawn from unique pairs in a given SOM cell, producing $n \choose 2$ points for a cell with $n$ spectroscopic galaxies. To fit the linear model, we take slices in $\Delta \rm{MAG\_GAAP\_Z}$ and fit each histogram of points to a normal distribution with a median and width, $\sigma$. Tails in the distribution cause deviations from the fitted normal that become larger with increasing |$\Delta M$|, which may lead to slight underestimation of the slope uncertainty. The fitting process is demonstrated visually in Fig. \ref{fig:appdzdm}. The medians at the bin midpoints are then fit with a linear regression that produces our estimate of $dz/dm$ for a given sample. 

For Fig. \ref{fig:dzdm_cell}, where there are fewer data points for each cell than for the entire sample, only two slices in magnitude were chosen, where they were chosen to jointly span $[\rm{min}(\Delta M_{\rm{cell}}),\rm{max}(\Delta M_{\rm{cell}})]$.

%%%%%%%%%%%%%%%%%%%%%%%%%%%%%%%%%%%%%%%%%%%%%%%%%%

% Don't change these lines
\bsp	% typesetting comment
\label{lastpage}
\end{document}